\documentclass[aps,prl,twocolumn,letterpaper,superscriptaddress]{revtex4-2}
\usepackage{graphicx}
\usepackage{CJK}
\usepackage{verbatim}
\usepackage{physics}
\usepackage[colorlinks,linkcolor=blue,anchorcolor=blue, citecolor=blue,urlcolor=blue,]{hyperref}
\usepackage{lineno}
\usepackage{bm}
\usepackage{mathptmx}
\usepackage{xcolor}
\makeatletter

\begin{document}
\begin{CJK*}{UTF8}{bsmi}
\title{Bogoliubov quasi-particles in superconductors are integer-charged particles inapplicable for non-Abelian quantum information braiding}
\author{Zhiyu Fan (\CJKfamily{gbsn}樊知宇)}
\affiliation{School of Physics and Astronomy, Shanghai Jiao Tong University, Shanghai 200240, China}
\author{Wei Ku (\CJKfamily{bsmi}顧威)}
\altaffiliation{corresponding email: weiku@sjtu.edu.cn}
\affiliation{School of Physics and Astronomy, Shanghai Jiao Tong University, Shanghai 200240, China}
\affiliation{Key Laboratory of Artificial Structures and Quantum Control (Ministry of Education), Shanghai 200240, China} 

\date{\today}

\begin{abstract}
We present a rigorous proof that under a number-conserving Hamiltonian, one-body quasi-particles generally possess quantized charge and inertial mass identical to the bare particles.
It follows that, Bogoliubov zero modes in the vortex of superconductors \textit{cannot} be their own anti-particles capable of braiding quantum information.
As such, the heavily pursued Majorana zero mode-based route for quantum computation requires a serious re-consideration.
This study further reveals the conceptual challenge in preparing and manipulating braidable quantum states via physical thermalization or slow external fields.
These profound results should reignite the long-standing quest for a number-conserving U(1) symmetric theory of superconductivity.

\end{abstract}

\maketitle
\end{CJK*}

\textit{Introduction} -
Computation has became an inseparable part of modern society.
As the classical computers approaching their quantum physical limit~\cite{powell2008}, significant amount of efforts have been made to seek the next-generation routes to push the computation capabilities.
Quantum computation~\cite{preskill2023quantumcomputing40years} has since become a intensive research field with profound technical the scientific challenges.

One of the promising routes being heavily pursued in the past decades is the use of Bogoliubov zero modes~\cite{caroli_bound_1964, gygi_electronic_1990} in the vortex core of topological superconductors~\cite{nayakNonAbelian2008, beenakker_search_2013}.
The theoretical justification is based on the current description of a Bogoliubov quasi-particle~\cite{gennes_self-consistent_1999} being a coherent superposition of particle and its antiparticle (``hole'').
As such, its zero mode, having exactly equal mixture of particle and hole, is its own anti-particle, namely a Majorana zero mode~\cite{read_paired_2000, kitaev_unpaired_2001}.
It follows that such zero modes possess non-Abelian statistics in two-dimension~\cite{read_paired_2000, ivanov_non-abelian_2001}, suitable for braiding many-body quantum information.
Together with their spatial non-locality that help resist decoherence against local perturbation, these Bogoliubov zero modes thus offers one of the most promising route towards industrializable quantum computation.

The theoretical foundation of this route is, however, intimately tied to the mainstream description of superconductivity via an effective quantum field with spontaneously broken U(1) symmetry, an aspect that has been long \textit{questioned} by masters of physics~\cite{Leggett_2001_SSB_SF} due to its violation of the profound particle conservation.
Correspondingly, whether Bogoliubov quasi-particles are allowed to be a coherent mixture of electron and hole, as to produce non-Abelian statistics, has recently been questioned as well~\cite{ortizManyBody2014, wangNumberconserving2017, knappNumberconserving2020, lapaRigorous2020,lin2022some}.
Furthermore, the current effective field description only arrives from tracing out ($N\!>\!2$)-body fluctuation/information.
Therefore, reassembling the resulting one-body zero modes back into a many-body system and utilizing the resulting non-Abelian statistics [equivalent to adding ($N\!>\!2$)-body quantum information \textit{by hand}] is hardly a reliable/legal practice at the conceptual level for many-body quantum information.

Here, we present a rigorous mathematical proof that under a number-conserving Hamiltonian, \textit{generally} one-body quasi-particles possess quantized charge and inertial mass identical to the bare particles.
Therefore, Bogoliubov zero modes, being the one-body quasi-particles in a superconductor, cannot possibly be their own anti-particles.
That is, these zero modes are not Majorana in nature and their statistics follows the standard Abelian fermionic statistics.
Therefore, a direct application of them in braiding many-body quantum information, as currently being heavily pursued in the field, is conceptually unfounded.
Furthermore, this study also reveals a conceptual challenge in preparing and manipulating braidable quantum states via physical thermalization or weak external fields.
These profound results should reignite the long-standing quest for a proper number-conserving U(1) symmetric theory of superconductivity.

\begin{figure}
    \hspace{-2.3cm}
    \includegraphics[scale=0.16]{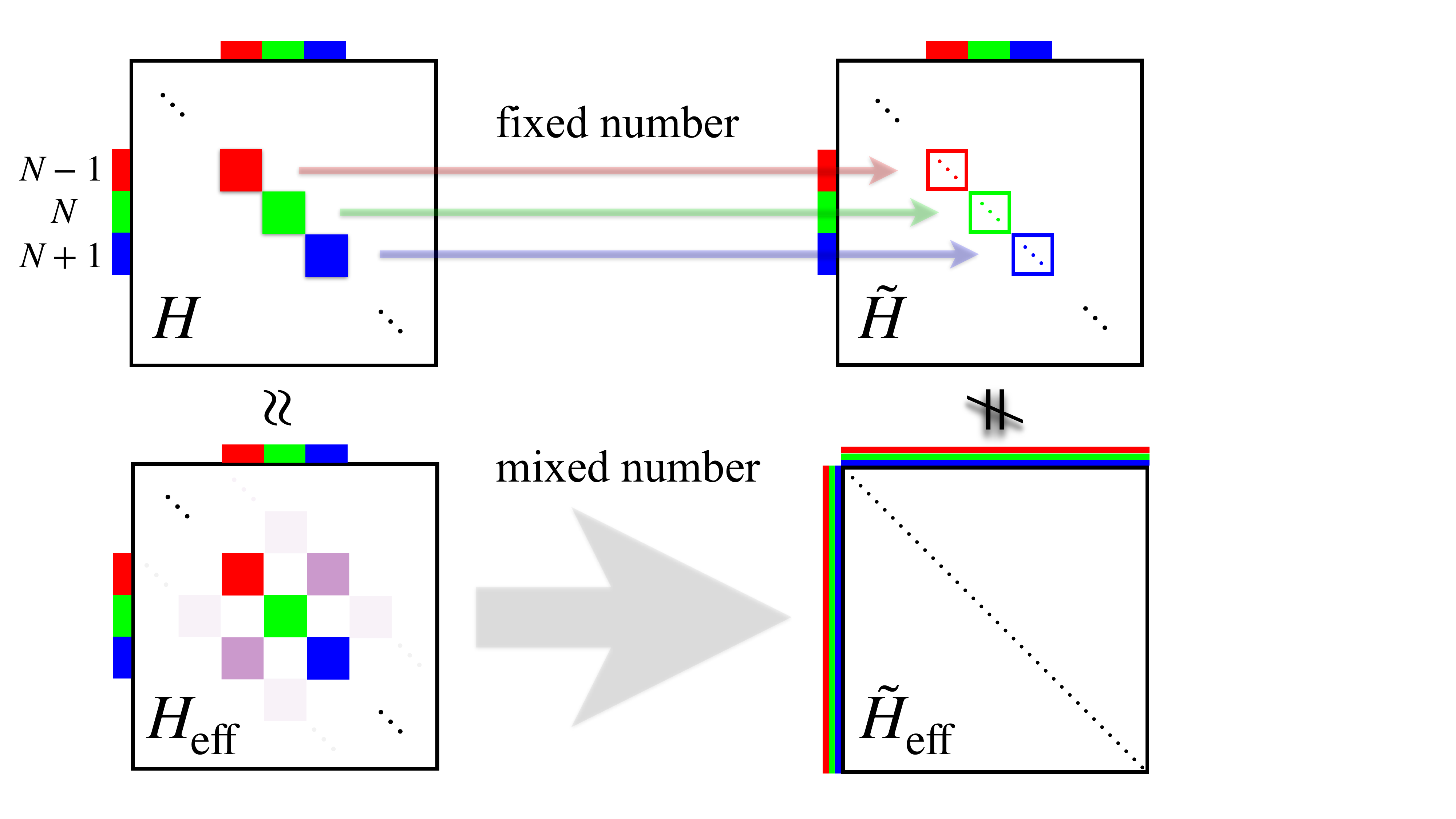}
    \hspace{-2cm}
    \caption{Matrix representation of a number-conserving many-body Hamiltonian $H$, with elements $\bra{I}H\ket{I^\prime}$ in a general basis $\ket{I}$ of Fock space, showing \textit{independent} diagonalization of each sector of fixed particle numbers in $\tilde{H}$ that results in conservation of $N$ in eigenstates.
    In contrast, $H_\mathrm{eff}$ for standard (effective field) theories of superconductivity contains approximate (problematic) couplings (purple blocks) between sectors of different particle numbers, resulting in \textit{artificial} coherence in the eigenstates beyond the physical dynamics described by $H$.}
    \vspace{-0.4cm}
    \label{fig1}
\end{figure}

\textit{Necessity of strict number conservation} -
Let's first recall the long-standing issue~\cite{Leggett_2001_SSB_SF, richardsonExact1964, bardeenGauge1957,nambu_quasi-particles_1960, baym_conservation_1961} in the standard theory of superconductivity concerning its violation of number conservation.
In contrast to particle physics, dynamics of electrons in condensed matter systems is too low in energy to allow number fluctuation.
Thus, generally all sensible description of electrons in condensed matter systems, for example,
\begin{align}    
    H = \sum_{ll^\prime} t_{ll^\prime} c_l^\dagger c_{l^\prime} +\frac{1}{2}\sum_{l_1 l_2 l_1^\prime l_2^\prime}U_{l_1 l_2,l_1^\prime l_2^\prime}c_{l_1}^\dagger c_{l_2}^\dagger c_{l_2^\prime}c_{l_1^\prime},
\label{generic_H}
\end{align}
contain a strict number conservation, $[N_c,H]=0$, where $N_c\equiv\sum_l c^\dagger_l c_l$ counts the total number of bare particles created by $c_l^\dagger$ and $l$ indexes the one-body Hilbert space spanned by atomic site, orbital, and spin degrees of freedom.
Consequently, as illustrated in Fig.~\ref{fig1}, eigenstates of $H$,
\begin{align}    
    \tilde{\ket{I}}^{(N)} \equiv \mathcal{U}^\dagger \ket{I^\prime}^{(N)} = \sum_{I^\prime} \ket{I^\prime}^{(N)} M_{I^\prime I}^{(N)},
\label{eigen_superposition}
\end{align}
only contains superposition (with coefficient $M_{I^\prime I}$) of the basis states, $\ket{I^\prime}$, within the subspace with the \textit{same} particle number $N$.
In other words, the unitary transformation $\mathcal{U}^\dagger$ respects the particle number as well,
\begin{equation}
  [N_c,\mathcal{U}^\dagger]=0.
  \label{N_c_U_commute}
\end{equation}

Such a strict number conservation in the quantum dynamical processes in $H$ implies that even in thermal ensembles using energy-based probability $\rho(H)$, for example the thermal average, $\langle O \rangle = \mathrm{Tr}( \rho O)=\sum_I \bra{I}\rho O \ket{I}$, of a physical observable, 
$O$, 
each sampled quantum state $\ket{I}$ must have strict number conservation as well.
In turn,
number-impaired observables such as $\langle c_l^\dagger c_{l^\prime}^\dagger \rangle$, must be \textit{strictly zero} in exact $N$-body treatments.

In great contrast to the above mathematically rigorous condition, due to introduction of \textit{approximate} number fluctuating dynamics, e.g. quadratic $c^\dagger c^\dagger$ terms, the standard descriptions of superconductivity~\cite{bardeenTheory1957, bogoljubovnew1958} instead result in ground states containing \textit{coherent} superpositions of states with a wide range of particle numbers $n$ around an average particle number $N$,
\begin{align}    
    \tilde{\ket{J}}_N = \sum_n B_N^n\tilde{\ket{J}}^{(n)} \equiv \sum_n B_N^n\Bigl[\sum_{J^\prime} \ket{J^\prime}^{(n)} M_{J^\prime J}^{(n)}\Bigr],
\label{BCS_superposition}
\end{align}
via coefficient $B_N^n$.
Furthermore, since such superposition has no physical origin from $H$, the energy expectation value,
\begin{align}    
    \langle H \rangle = \sum_n |B_N^n|^2 \langle H \rangle^{(n)}\equiv \sum_n |B_N^n|^2~~ ^{(n)}\tilde{\bra{J}} H \tilde{\ket{J}}^{(n)},
\label{BCS_energy}
\end{align}
is naturally insensitive to the relative phase of $B_N^n$ between sectors with different particle numbers $n$, thus a U(1) ``symmetry''.
Upon picking an arbitrary relative phase $\theta$, the superconducting ground state is therefore conveniently (but strictly speaking incorrectly) regarded as one with spontaneously broken U(1) symmetry.
Correspondingly, $\Delta_{ll^\prime}\equiv\langle c_l^\dagger c_{l^\prime}^\dagger \rangle$ becomes non-zero and serves as an ``order parameter''.

Such an effective field description has been quite successful in describing many unique one-body and two-body properties of superconductivity, \textit{when} the property of interest is insensitive to particle conservation.
Examples include one-body spectral function~\cite{Richard2015ARPESMO, PhysRevB.79.020505, RevModPhys.75.473}, tunneling~\cite{RevModPhys.79.353, leespectroscopic2009}, and Andreev reflection~\cite{RevModPhys.77.109}, to name a few.

However, properties sensitive to particle conservation are well-known to be highly problematic.
To give a representative example of this long-standing and rather serious issue, theoretical optical conductivities would generally \textit{fail}~\cite{bardeenGauge1957} to uphold the sum-rule without specially designed fixes~\cite{nambu_quasi-particles_1960, baym_conservation_1961}, even though it should have been \textit{automatically} fulfilled according to the continuity equation for physical transport processes.
Given that even two-body properties already irreparably suffer from the disrespect of particle conservation, the validity of braiding-associated $N$-body quantum information involving \textit{multiple} one-body zero modes, is thus seriously questionable~\cite{ortizManyBody2014}.

Some existing literature seemed to imply that systems with spontaneously broken symmetry are exceptional to the above consideration of number conservation.
This is, however, against the dynamics of quantum state thermalization (c.f. \cite{supplementary}).
Specifically, \textit{physical} states with spontaneous broken symmetry actually result from \textit{incomplete} quantum thermalization limited by the finite time scales of experiments~\cite{MoreAnderson1972} (with slower off-diagonal fluctuation not active yet).
Therefore, the above rigorous consideration on number conserving dynamics and all its implications must still apply.

\textit{One-body quasi-particles} -
In this study, we focus on the rigorous implications of particle conservation on one-body quasi-particles, as typically observed (experimentally or theoretically) through peaks in the one-body Green's function, $\langle G_l(\omega)\rangle$, with well-defined widths in energy $\omega$ and momentum.
Here $G_l(\omega)$ is the Fourier transform of the time-ordered Green's function operator~\cite{Mattuck1976AGT}, $G_l(t) \equiv -i\mathcal{T} \Bigl[ c_l(t)c_l^\dagger(0) \Bigr]$,
under a time ordering operator $\mathcal{T}$.

In standard effective field theories, these one-body quasi-particles, for example the Bogoliubov quasi-particles in superconductors, typically emerge as the dressed particles $\gamma_l^\dagger$ that absorb the dominant quantum dynamics of short time scales, such that their slower dynamics can be approximately described by a \textit{diagonal} low-energy effective Hamiltonian~\cite{Mattuck1976AGT},
\begin{equation}
\label{H_diag_QP}
H^\mathrm{eff}=\sum_l E_l \gamma_l^\dagger \gamma_l.
\end{equation}
Correspondingly, eigenstates $\ket{\tilde{J}}$ (within an approximate subspace) can be accessed by adding some number of one-body quasi-particles $\gamma_l^\dagger$ to a known eigenstate (for example, the ground state $\ket{\tilde{J}_0}$)~\cite{andersonconcepts1997},
\begin{equation}
    \ket{\tilde{J}_{\{l\},\tilde{J}_0}}\approx\prod_{l^\prime\in\{l\}} \gamma_{l^\prime}^\dagger \ket{\tilde{J}_0}.
\label{apprx_eigenstate}
\end{equation}

\textit{Eigen-particles as one-body quasi-particles with full $N$-body information} -
In exact $N$-body treatments, such dressed particles that \textit{fully} absorb $N$-body quantum dynamics of short time scales are known as ``eigen-particles''~\cite{whiteNumerical2002, kanno_method_1969,heggUniversal2024}, as obtained through a unitary transformation of bare particles,
\begin{equation}
    \tilde{c}^{\dagger}_l\equiv \mathcal{U}^{\dagger} c_l^{\dagger}\mathcal{U},
\label{eigen-p}
\end{equation}
within the $N$-body Hilbert space (as in the Schrieffer-Wolff transformation~\cite{SW_transform}) that leaves only \textit{diagonal} processes in $H$,
\begin{align}
    \tilde{H}[\{\tilde{c}^{\dagger}_l\}] &\equiv H[\{c^{\dagger}_l\}] = \mathcal{U}H[\{\tilde{c}^{\dagger}_l\}]\mathcal{U}^{\dagger} \nonumber \\ 
    &=\sum_{l_1} \epsilon_{l_1} \tilde{c}^{\dagger}_{l_1} \tilde{c}_{l_1} + \frac{1}{2!} \sum_{l_1 l_2}  \epsilon_{l_1 l_2} \tilde{c}^{\dagger}_{l_1} \tilde{c}^{\dagger}_{l_2} \tilde{c}_{l_2} \tilde{c}_{l_1} \nonumber \\
    &+ \frac{1}{3!} \sum_{l_1 l_2 l_3}  \epsilon_{l_1 l_2 l_3} \tilde{c}^{\dagger}_{l_1} \tilde{c}^{\dagger}_{l_2} \tilde{c}^{\dagger}_{l_3} \tilde{c}_{l_3} \tilde{c}_{l_2} \tilde{c}_{l_1} + \cdots,
\label{H_eigen}
\end{align}
analogous to Eq.~\ref{H_diag_QP}.
Correspondingly, their 1- to $N$-body densities are all constants of motion, and product states of eigen-particles, with respect to the zero-electron true vacuum, $\ket{0}$,
\begin{equation}
    \ket{\tilde{I}_{\{l\}}}=\prod_{l^\prime\in\{l\}} \tilde{c}_{l^\prime}^\dagger \ket{0},
\label{exact_eigenstate}
\end{equation}
\textit{rigorously} form a complete set of orthonormal eigenstates of $H$~\cite{supplementary}.
With respect to any eigenstate $\ket{\tilde{I}_0}\equiv\ket{\tilde{I}_{\{l_0\}}}$, this implies that product states containing additional eigen-particles, $\ket{\tilde{I}_{\{l\},I_0}}\equiv\prod_{l^\prime\in\{l\}} \tilde{c}_{l^\prime}^\dagger \ket{\tilde{I}_0}$, are eigenstates as well, perfectly reproducing the property of one-body quasi-particles in Eq.~\ref{apprx_eigenstate}.

As a matter of fact,
$G_l(\omega)$ represented in eigen-particles,
\begin{align}
    G_l(\omega) &= c_l^\dagger \Bigl[(\omega - E_l-i0^+)^{-1} \tilde{c}_l  \Bigr] + \Bigl[
    (\omega - E_l+i0^+)^{-1} \tilde{c}_l \Bigr] c_l^\dagger \nonumber \\
    &+ c_l^{\dagger}\Bigl[\frac{1}{2!}\sum_{l_{1}l_{2}l_{1}'}(\omega-E_{l_{1}l_{2};l_{1'}}-i0^{+})^{-1}A_{l_{1}l_{2};ll_{1}'}^{*}\tilde{c}_{l_{1}'}^{\dagger}\tilde{c}_{l_{2}}\tilde{c}_{l_{1}}\Bigr] \nonumber \\
    &+ \Bigl[\frac{1}{2!}\sum_{l_{1}l_{2}l_{1}'}(\omega-E_{l_{1}l_{2};l_{1'}}+i0^{+})^{-1}A_{l_{1}l_{2};ll_{1}'}^{*}\tilde{c}_{l_{1}'}^{\dagger}\tilde{c}_{l_{2}}\tilde{c}_{l_{1}} \Bigr]c_l^{\dagger}\nonumber \\
    &+\cdots,
    \label{G_in_eigen}
\end{align}
with coefficients $A_{l_1 l_2;ll_1^\prime}$ obtained from,
\begin{equation}
c_l^\dagger=\mathcal{U}\tilde{c}_l^\dagger\mathcal{U}^\dagger = \tilde{c}_l^\dagger + \frac{1}{2!}\sum_{l_1 l_2 l_1^\prime} A_{l_1 l_2;l l_1^\prime}\tilde{c}_{l_1}^\dagger\tilde{c}_{l_2}^\dagger  \tilde{c}_{l_1^\prime}+\cdots
\end{equation}
shows that energies of well-defined quasi-particle peaks (that dominate experimental or theoretical spectra) are given in the first line of Eq.~\ref{G_in_eigen} by the one-body ``eigen-particle energies'', $E_l$, defined through,
$\tilde{c}_l^\dagger E_l \equiv [\tilde{H},\tilde{c}_l^\dagger]$.
The rest of Eq.~\ref{G_in_eigen}, containing \textit{summation} over energies of $(n+1; n)$-body (particle; hole) fluctuations defined through $\tilde{c}^\dagger_{l_1}\tilde{c}^\dagger_{l_2}\cdots\tilde{c}_{l_1^\prime} E_{l_1l_2\cdots;\cdots l_1^\prime} \equiv [\tilde{H},\tilde{c}^\dagger_{l_1}\tilde{c}^\dagger_{l_2}\cdots\tilde{c}_{l_1^\prime}]$, produces the observed continuum, peak broadening, and strong correlation-constrained one-body quasi-particles~\cite{supplementary}.

Therefore, while typically observed experimentally through bare-particles, one-body quasi-particles are \textit{rigorously} described as eigen-particles in exact $N$-particle treatments.
The typical field theoretical descriptions of the same only capture the \textit{few-body reduced} information sufficient for some one- and two-body properties.
Reliable quantum information of the system, however, would require the full $N$-body information as encoded in eigen-particles.

\textit{Eigen-particles carry bare charge and inertial mass} -
Having established this correspondence, let's examine the rigorous implications of number conservation of $H$ on eigen-particles.\\ \\
\textit{Theorem 1:}\\
Under a number-conserving $H$, eigen-particles carry exactly \textit{one} particle number of bare particles.\\ \\
\textit{Proof:}\\
According to Eq.~\ref{N_c_U_commute}, $[N_c,\mathcal{U}^\dagger]=0$,
\begin{align}
N_c\tilde{c}^\dagger_l - \tilde{c}^\dagger_l N_c &= [N_c,\tilde{c}^\dagger_l] = [N_c, \mathcal{U}^\dagger c^\dagger_l \mathcal{U}] = \mathcal{U}^\dagger [N_c, c^\dagger_l] \mathcal{U} \nonumber \\
&= \mathcal{U}^\dagger c^\dagger_l \mathcal{U} = \tilde{c}_l^\dagger, 
\label{one_particle}
\end{align}
or $N_c\tilde{c}^\dagger_l=\tilde{c}^\dagger_l(N_c+1)$ , indicates that operator $\tilde{c}^\dagger_l$ increases the bare particle number $N_c$ by 1.

Immediately following this, the electric charge and inertial mass attached to each bare particle are therefore carried by eigen-particles as well.
(This conclusion does not rely on, nor does it imply, locality of the charge and inertial mass.
See the discussion below Eq.~\ref{eig-p-bcs} for Bogoliubov quasi-particles.)

\textit{Proper Bogoliubov quasi-particles} -
Experts in superconductivity might be immediately alerted by the above rigorous conclusion, as it obviously invalidates the well-accepted nature of Bogoliubov quasi-particles~\cite{bogoljubovnew1958},
\begin{equation}
    \begin{cases}
    \gamma_{k\uparrow}^{\dagger}=u_{k}\textcolor{blue}{c_{k\uparrow}^{\dagger}}-v_{k}\textcolor{red}{c_{-k\downarrow}}\\
    \gamma_{-k\downarrow}=u_{k}\textcolor{blue}{c_{-k\downarrow}}+v_{k}\textcolor{red}{c_{k\uparrow}^{\dagger}}
    \end{cases}
    \label{Bogoliubov_QP}
\end{equation}
being a coherent superposition of electrons and holes, with probability amplitudes $u_k$ and $v_k$,
here $k$ denotes the momentum.
Indeed, $[N_c,\gamma_{k\sigma}^\dagger]\neq 1$ violates the rigorous requirement in Eq.~\ref{one_particle}.
Therefore, $\gamma_{k\sigma}^\dagger$ in Eq.~\ref{Bogoliubov_QP} should only be considered as a (problematic) approximate reduction of the eigen-particles, justifiable \textit{only} under unphysical number-changing dynamics, $c_{k'\uparrow}^\dagger c_{-k'\downarrow}^\dagger$.
Correspondingly, their well-accepted nature of mixed electron and hole should be regarded as an \textit{artifact} from the violation of particle conservation, in association with the employment of an effective field under a spontaneously broken U(1) symmetry.
A ``proper'' description of Bogoliubov quasi-particles through eigen-particles (or their reduced counterparts) must adhere to Eq.~\ref{one_particle} instead and preserve the quantized charge and inertial mass of bare electrons.

As an illustration to bring back some familiarity, let's adapt in $H$ BCS's original proposal of electron pairing due to an emergent ``relevant'' attractive interaction of the form~\cite{bardeenTheory1957},
\begin{align}
    H_\mathrm{BCS}=-\frac{V}{\Omega}\sum_{k k^\prime} c^{\dagger}_{k\uparrow}c^{\dagger}_{-k\downarrow } c_{-k^\prime\downarrow}c_{k^\prime\uparrow},
\label{H_attraction}
\end{align}
of strength $V$, where $\Omega$ denotes system size. 
The corresponding eigen-particles would then be dominated by,
\begin{align}
   \textcolor{blue}{\tilde{c}_{k\sigma}^\dagger} &\approx \textcolor{blue}{c_{k\sigma}^\dagger}
        +\sum_{k_{1}}\tilde{A}^{k\sigma}_{k_1} \textcolor{brown}{c_{k_{1}\uparrow}^\dagger c_{-k_{1}\downarrow}^\dagger} \textcolor{red}{c_{-k\bar{\sigma}}}\nonumber \\
\label{eig-p-bcs}
        &+ \sum_{k_1 k_1^\prime} \tilde{A}^{k\sigma}_{k_1; k_1^\prime} \textcolor{blue}{c_{k\sigma}^\dagger} c_{k_1\uparrow}^\dagger c_{-k_1\downarrow}^\dagger c_{-k_1^\prime\downarrow}c_{k_1^\prime\uparrow} \\
        &+ \sum_{k_1 k_2 k_1^\prime} \tilde{A}^{k\sigma}_{k_1 k_2; k_1^\prime} \textcolor{brown}{c_{k_1\uparrow}^\dagger c_{-k_1\downarrow}^\dagger c_{k_2\uparrow}^\dagger c_{-k_2\downarrow}^\dagger c_{-k_1^\prime\downarrow} c_{k_1^\prime\uparrow}} \textcolor{red}{c_{-k\bar{\sigma}}}
        +\cdots, \nonumber
\end{align}        
with coefficients $\tilde{A}$'s determined from the diagonalization of $H$ under the canonical condition,
\begin{align}
    \{\tilde{c}_{k\sigma}, \tilde{c}_{k^\prime \sigma^\prime}^\dagger\} &= \mathcal{U}^\dagger\{c_{k\sigma}, c^{\dagger}_{k^\prime\sigma^\prime}\}\mathcal{U} = \delta_{kk^\prime}\delta_{\sigma\sigma^\prime},\nonumber\\
    \{\tilde{c}_{k\sigma}, \tilde{c}_{k^\prime \sigma^\prime}\} &= \mathcal{U}^\dagger\{c_{k\sigma}, c_{k^\prime\sigma^\prime}\}\mathcal{U} =0.
\end{align}

One might now recognize the familiar coherent superposition between \textcolor{blue}{$c_{k\sigma}^\dagger$} (in blue) and \textcolor{red}{$c_{-k\bar{\sigma}}$} (in red), except that they are accompanied by additional creation and annihilation operators to strictly satisfy Eq.~\ref{one_particle}.
Upon accepting an number-impaired ``improper'' mean-field, $-\frac{V}{\Omega}  \sum_{k'} \langle c_{k'\uparrow}^\dagger c_{-k'\downarrow}^\dagger \rangle$, it is straightforward~\cite{supplementary} to reduce such proper Bogoliubov quasi-particles into the familiar form in Eq.~\ref{Bogoliubov_QP}.
Nonetheless, as illustrated above via Fig.~\ref{fig1},
such number-impaired meanfield must \textit{strictly vanish} under a number conserving Hamiltonian.
Therefore, despite the great success~\cite{bardeenTheory1957, bogoljubovnew1958} of improper Bogoliubov quasi-particles in Eq.~\ref{Bogoliubov_QP} in describing some one- and two-body properties, it is absolutely necessary to adhere to their proper number-conserving description, Eq.~\ref{eig-p-bcs}, when considering $N$-body quantum information.

The structure of Eq.~\ref{eig-p-bcs} also offers reconciliation with the current lore that the electric charge of low-energy improper Bogoliubov quasi-particles in Eq.~\ref{Bogoliubov_QP} can be completely screened (except a ``Z2'' charge to retain the AB phase)~\cite{KivelsonBogoliubov1990, hanssonsuperconductors2004}.
Notice, however, that such improper description misses the components in brown in Eq.~\ref{eig-p-bcs} that contain the complementary contributions to the quantized charge and inertial mass, so an incomplete account for the total charge and inertial mass is expected.
Of course, real experimental observations of Bogoliubov quasi-particles, through either photoemission~\cite{PhysRevB.79.020505} or tunneling~\cite{leespectroscopic2009}, \textit{unambiguously} add the \textit{fully quantized} charge and inertial mass to the sample, in perfect agreement with the proper $N$-body description.

\textit{Inability to braid quantum information} -
Given the strict number conservation in $H$, it follows that generally electronic eigen-particles, such as the proper Bogoliubov quasi-particles, cannot possibly be their own anti-particles, as intuitively confirmed by the non-self-adjoint form of Eq.~\ref{eig-p-bcs}.\\ \\
\textit{Theorem 2:}\\
Under a number-conserving $H$, eigen-particles cannot be Majorana particles.\\ \\
\textit{Proof:}\\
Assuming one-body eigen-particles can be their anti-particles, $\tilde{c}_l^\dagger=\tilde{c}_l$,
\begin{equation}
\tilde{c}_l^\dagger=[N_c, \tilde{c}_l^\dagger]=-[\tilde{c}_l^\dagger,N_c]=-[\tilde{c}_l,N_c]=-\tilde{c}_l=-\tilde{c}_l^\dagger,
\end{equation}
can only be satisfied by $\tilde{c}_l^\dagger=0$, thus establishing the theorem.
(Here, $\tilde{c}_l = [\tilde{c}_l,N_c]$ is implied by Eq.~\ref{one_particle}.)\\

Now, since proper Bogoliubov quasi-particles are in fact massive integer-charged particles, rather than Majorana particles, absence of the necessary property, $\tilde{c}_l^\dagger=\tilde{c}_l$, inevitably invalidates the theoretical foundation~\cite{ivanov_non-abelian_2001} for the current proposals of non-Abelian quantum information braiding directly using Bogoliubov zero modes.
In other words, concerning the \textit{coherent} many-body states of the \textit{entire} sample, there only exist charge-$e$ dressed electrons (or holes) in the vortex core of superconductors, but no Majorana zero mode to facilitate the proposed non-Abelian braiding.
Essentially, as clearly illustrated from the distinction between Eq.~\ref{Bogoliubov_QP} and \ref{eig-p-bcs}, the common misidentification of Bogoliubov zero modes as Majorana ones in standard superconducting theories is rooted in their serious violation of the strict particle conservation of electron dynamics in condensed matter systems.

\textit{Alternative proof via eigenstate structure} -
Another way to observe Bogoliubov zero modes' inability to support non-Abelian braiding, independent of specific particle properties (i.e charge locality or long-range screening), is through the rigorous \textit{product states} structure, Eq.~\ref{exact_eigenstate}, of physical eigenstates in the eigen-particle representation.
Since each Bogoliubov zero modes are pined to the center of a vortex, the degenerate ground states of the system can be labeled by eigen-particles' occupation of these spatially distinguishable vortices, $\ket{1,1,0,1,0,\dots}$.
Therefore, a braiding operation, $B_{ii^\prime}$ between vortices at locations $i$ and $i^\prime$ (that eventually returns them back to their original locations) would simply result in the original product state (that is, $B_{ii^\prime}=1$) instead of another eigenstate within the degenerate manifold as required by non-Abelian braiding.
Consistently, given $M$ eigen-particles occupying $L$ vortices, the size of the degenerate subspace is therefore $C^L_M$, rather than the $2^{L/2}$ (or $2^{L/2-1}$ for each parity)
in current Majorana proposals~\cite{ivanov_non-abelian_2001}.
Such a significant difference should be detectable through, for example, heat capacity experiments or properly scaled finite-size computations.

Beside the standard theories with broken U(1) symmetry via long-range ordering of local order parameters, $\Delta_i\equiv\langle c^\dagger_{i\uparrow} c^\dagger_{i\downarrow}\rangle$, at each site $i$,
some superconductivity studies~\cite{giamarchi2003quantum, NumberSau2011, AttractiveWhite1989, SuperconductivityLuk1990, HoleWhite1997} did uphold (or impose) number conservation.
Having no measurable local order parameter, $\Delta_i=0$, superconductivity in these studies is typically detected instead via the two-body density matrix, $\rho^{(2)}\equiv\langle c^\dagger_{i\uparrow}c^\dagger_{i\downarrow}c_{i'\downarrow}c_{i'\uparrow}\rangle$, which signals the emergence of bound bi-electrons and their long-range propagation, two of the essential ingredients for superconductivity.
Adapting a long-range order picture (via the dominant eigenvector of $\rho^{(2)}$) or not, these studies would result in unbraidable eigenstates, as the number conservation \textit{alone} would invalidate Eq.~\ref{Bogoliubov_QP} in these studies according to theorems 1 and 2.

\textit{Braidable states and challenge in their preparation and manipulation} -
Generally, the above discussion rigorously invalidates all possibilities in \textit{directly} braiding one-body quasi-particles in number-conserving systems.
It does not, however, rule out scenario of braiding objects that can be engineered in eigenstates made from extensive superposition of \textit{degenerate} product states of eigen-particles~\cite{mooreNonabelions1991}.

Nonetheless, the above discussion on eigen-particles still indicates a generic challenge in preparing and manipulating such braidable states in materials.
Recall that it is the physical \textit{dynamical} processes in $H$ that is behind the eigenstates thermalization~\cite{DeutschETH1991,SrednickiETH2012} of real materials.
Similarly, it is the \textit{integer}-quantized electrons that the external fields directly couple to during weak non-destructive external manipulation.
Given integer-quantized eigen-particles' ability to fully absorb quantum dynamics in $H$, slow physical processes in preparation and manipulation would, by default, \textit{only} reach product states of eigen-particles.

In contrast, states made of extensive superposition of these product states risk imposing \textit{unphysical} coherence beyond the dynamics encoded in $H$.
Therefore, as speculated from the very beginning~\cite{mooreNonabelions1991}, preparation and manipulation of such braidable states would be physically challenging to achieve, even at the conceptual level.
In fact, this is exactly analogous to the lack of experimental observation of moment-less ferromagnetic states, $\frac{1}{\sqrt{2}}(\ket{\uparrow}+\ket{\downarrow})$, in macroscopic materials, as opposed to the commonly observed product states with total magnetic moment, $\ket{\uparrow}~\equiv\prod_l \tilde{c}_{l\uparrow}^\dagger\ket{0}$.
Such moment-less states contain extra coherence beyond the physical dynamics in $H$, and thus challenging to prepare in thermal equilibrium.

To give a trivial illustration on this generic issue of artificial coherence associated with degeneracy in quantum mechanics, consider an array of centimeter spaced lattice of optical traps having a single atom without physically meaningful inter-trap couplings.
In such a spatially decoupled system, an atom in the local ground state of a single trap at position $x$, $a_x^\dagger$, is already an eigen-particle of the system.
Correspondingly, the system hosts a high level of degeneracy associated with the freedom in $x$.
Obviously, it is physically achievable to prepare and manipulate the trivial product states $a_x^\dagger \ket{0}$.
In contrast, states made of \textit{artificial} superposition of these degenerate eigenstates, 
$\tilde{a}_k^\dagger \equiv \sum_x^M \tilde{a}_x^\dagger e^{ikx}/\sqrt{M}$,
would be extremely challenging to prepare or manipulate, given no physical processes in $H$ to establish the extra coherence.

Following the above consideration, perhaps a more achievable (but hardly explored) route for engineering braidable systems is through explicitly breaking number conservation of a system.
This can be achieved by driving the systems with significant amount of charge fluctuation, or by strongly coupling the systems to a well-designed environment with efficient exchange of electrons.

\textit{Toward a number-conserving theory of superconductivity} -
As for the superconducting ground state, the above concern on degeneracy-enabled artificial coherence equally applies.
Indeed, while one is mathematically allowed to introduce artificial superposition of (nearly) degenerate ground states of $H-\mu N$ from sectors of different particle numbers, the resulting U(1) symmetry broken states have absolutely no dynamical origin from $H$ (c.f. Fig.~\ref{fig1}) and thus would not result from regular thermal equilibrium.

All significant implications of number conservation above should remind experts of the long-standing concerns~\cite{Leggett_2001_SSB_SF,richardsonExact1964, bardeenGauge1957,nambu_quasi-particles_1960, baym_conservation_1961} on the spontaneously broken U(1) symmetry in effective field descriptions of superconductivity.
As clearly discussed around Fig~\ref{fig1}, such description violates the profound number-conservation and must be viewed as a (highly useful) approximation risking unphysical properties, such as failing to automatically uphold the physical sum rule in dynamical transports, or mistaking Bogoliubov zero modes as Majorana ones.
Our study should thus reignite a serious reconsideration of number conservation toward proper U(1) \textit{symmetric} theories of superconductivity.
Particularly, the general eigen-particle representation employed here is likely to offer a convenient representation that explicitly encodes the essential many-body dynamics with strict number conservation~\cite{heggUniversal2024}.

\textit{Conclusion} -
In short, we present a mathematical proof that under a number-conserving Hamiltonian, eigen-particles as the $N$-body representation of the one-body quasi-particles, rigorously contains particle number 1 of bare particles.
They thus possess the same quantized charge and inertial mass of the bare electrons, and can never be their own anti-particles.
This result corrects a long-standing misconception on experimentally observed Bogoliubov quasi-particles being a mixture of electrons and holes, as resulting from approximate effective field theories.
Correspondingly, their zero modes are not Majorana ones and cannot be used for braiding $N$-body quantum information.
Furthermore, this study reveals a conceptual challenge in preparing and manipulating braidable quantum states via physical thermal equilibrium or slow external fields.
These profound results should reignite a long-standing quest for a number-conserving, U(1) \textit{symmetric} theory of superconductivity.

\begin{acknowledgments}
We thank Xiang Li, Hans T. Hansson, Tao Zeng, Matteo Baggioli, and Tsutomu Yanagida for useful discussions. This work is supported by National Natural Science Foundation of China (NSFC) under Grants No. 12274287 and No.12042507, and Innovation Program for Quantum Science and Technology No.2021ZD0301900.
\end{acknowledgments}

\bibliography{main_v3}

\begin{thebibliography}{51}%
\makeatletter
\providecommand \@ifxundefined [1]{%
 \@ifx{#1\undefined}
}%
\providecommand \@ifnum [1]{%
 \ifnum #1\expandafter \@firstoftwo
 \else \expandafter \@secondoftwo
 \fi
}%
\providecommand \@ifx [1]{%
 \ifx #1\expandafter \@firstoftwo
 \else \expandafter \@secondoftwo
 \fi
}%
\providecommand \natexlab [1]{#1}%
\providecommand \enquote  [1]{``#1''}%
\providecommand \bibnamefont  [1]{#1}%
\providecommand \bibfnamefont [1]{#1}%
\providecommand \citenamefont [1]{#1}%
\providecommand \href@noop [0]{\@secondoftwo}%
\providecommand \href [0]{\begingroup \@sanitize@url \@href}%
\providecommand \@href[1]{\@@startlink{#1}\@@href}%
\providecommand \@@href[1]{\endgroup#1\@@endlink}%
\providecommand \@sanitize@url [0]{\catcode `\\12\catcode `\$12\catcode
  `\&12\catcode `\#12\catcode `\^12\catcode `\_12\catcode `\%12\relax}%
\providecommand \@@startlink[1]{}%
\providecommand \@@endlink[0]{}%
\providecommand \url  [0]{\begingroup\@sanitize@url \@url }%
\providecommand \@url [1]{\endgroup\@href {#1}{\urlprefix }}%
\providecommand \urlprefix  [0]{URL }%
\providecommand \Eprint [0]{\href }%
\providecommand \doibase [0]{https://doi.org/}%
\providecommand \selectlanguage [0]{\@gobble}%
\providecommand \bibinfo  [0]{\@secondoftwo}%
\providecommand \bibfield  [0]{\@secondoftwo}%
\providecommand \translation [1]{[#1]}%
\providecommand \BibitemOpen [0]{}%
\providecommand \bibitemStop [0]{}%
\providecommand \bibitemNoStop [0]{.\EOS\space}%
\providecommand \EOS [0]{\spacefactor3000\relax}%
\providecommand \BibitemShut  [1]{\csname bibitem#1\endcsname}%
\let\auto@bib@innerbib\@empty
\bibitem [{\citenamefont {Powell}(2008)}]{powell2008}%
  \BibitemOpen
  \bibfield  {author} {\bibinfo {author} {\bibfnamefont {J.~R.}\ \bibnamefont
  {Powell}},\ }\bibfield  {title} {\bibinfo {title} {The quantum limit to
  moore's law},\ }\href@noop {} {\bibfield  {journal} {\bibinfo  {journal}
  {Proceedings of the IEEE}\ }\textbf {\bibinfo {volume} {96}},\ \bibinfo
  {pages} {1247} (\bibinfo {year} {2008})}\BibitemShut {NoStop}%
\bibitem [{\citenamefont
  {Preskill}(2023)}]{preskill2023quantumcomputing40years}%
  \BibitemOpen
  \bibfield  {author} {\bibinfo {author} {\bibfnamefont {J.}~\bibnamefont
  {Preskill}},\ }\href {https://arxiv.org/abs/2106.10522} {\bibinfo {title}
  {Quantum computing 40 years later}} (\bibinfo {year} {2023}),\ \Eprint
  {https://arxiv.org/abs/2106.10522} {arXiv:2106.10522 [quant-ph]} \BibitemShut
  {NoStop}%
\bibitem [{\citenamefont {Caroli}\ \emph {et~al.}()\citenamefont {Caroli},
  \citenamefont {De~Gennes},\ and\ \citenamefont
  {Matricon}}]{caroli_bound_1964}%
  \BibitemOpen
  \bibfield  {author} {\bibinfo {author} {\bibfnamefont {C.}~\bibnamefont
  {Caroli}}, \bibinfo {author} {\bibfnamefont {P.~G.}\ \bibnamefont
  {De~Gennes}},\ and\ \bibinfo {author} {\bibfnamefont {J.}~\bibnamefont
  {Matricon}},\ }\bibfield  {title} {\bibinfo {title} {Bound fermion states on
  a vortex line in a type {II} superconductor},\ }\href
  {https://doi.org/10.1016/0031-9163(64)90375-0} {\bibfield  {journal}
  {\bibinfo  {journal} {Physics Letters}\ }\textbf {\bibinfo {volume} {9}},\
  \bibinfo {pages} {307}}\BibitemShut {NoStop}%
\bibitem [{\citenamefont {Gygi}\ and\ \citenamefont
  {Schluter}()}]{gygi_electronic_1990}%
  \BibitemOpen
  \bibfield  {author} {\bibinfo {author} {\bibfnamefont {F.}~\bibnamefont
  {Gygi}}\ and\ \bibinfo {author} {\bibfnamefont {M.}~\bibnamefont
  {Schluter}},\ }\bibfield  {title} {\bibinfo {title} {Electronic tunneling
  into an isolated vortex in a clean type-{II} superconductor},\ }\href
  {https://doi.org/10.1103/PhysRevB.41.822} {\bibfield  {journal} {\bibinfo
  {journal} {Physical Review B}\ }\textbf {\bibinfo {volume} {41}},\ \bibinfo
  {pages} {822}}\BibitemShut {NoStop}%
\bibitem [{\citenamefont {Nayak}\ \emph {et~al.}(2008)\citenamefont {Nayak},
  \citenamefont {Simon}, \citenamefont {Stern}, \citenamefont {Freedman},\ and\
  \citenamefont {Das~Sarma}}]{nayakNonAbelian2008}%
  \BibitemOpen
  \bibfield  {author} {\bibinfo {author} {\bibfnamefont {C.}~\bibnamefont
  {Nayak}}, \bibinfo {author} {\bibfnamefont {S.~H.}\ \bibnamefont {Simon}},
  \bibinfo {author} {\bibfnamefont {A.}~\bibnamefont {Stern}}, \bibinfo
  {author} {\bibfnamefont {M.}~\bibnamefont {Freedman}},\ and\ \bibinfo
  {author} {\bibfnamefont {S.}~\bibnamefont {Das~Sarma}},\ }\bibfield  {title}
  {\bibinfo {title} {Non-{{Abelian}} anyons and topological quantum
  computation},\ }\href {https://doi.org/10.1103/RevModPhys.80.1083} {\bibfield
   {journal} {\bibinfo  {journal} {Reviews of Modern Physics}\ }\textbf
  {\bibinfo {volume} {80}},\ \bibinfo {pages} {1083} (\bibinfo {year}
  {2008})}\BibitemShut {NoStop}%
\bibitem [{\citenamefont {Beenakker}()}]{beenakker_search_2013}%
  \BibitemOpen
  \bibfield  {author} {\bibinfo {author} {\bibfnamefont {C.~W.~J.}\
  \bibnamefont {Beenakker}},\ }\bibfield  {title} {\bibinfo {title} {Search for
  majorana fermions in superconductors},\ }\href
  {https://www.annualreviews.org/content/journals/10.1146/annurev-conmatphys-030212-184337}
  {\bibfield  {journal} {\bibinfo  {journal} {Annual Review of Condensed Matter
  Physics}\ }\textbf {\bibinfo {volume} {4}},\ \bibinfo {pages}
  {113}}\BibitemShut {NoStop}%
\bibitem [{\citenamefont {Gennes}()}]{gennes_self-consistent_1999}%
  \BibitemOpen
  \bibfield  {author} {\bibinfo {author} {\bibfnamefont {P.~G.~D.}\
  \bibnamefont {Gennes}},\ }\bibfield  {title} {\bibinfo {title} {The
  self-consistent field method},\ }in\ \href@noop {} {\emph {\bibinfo
  {booktitle} {Superconductivity Of Metals And Alloys}}}\ (\bibinfo
  {publisher} {{CRC} Press})\ \bibinfo {note} {num Pages: 34}\BibitemShut
  {NoStop}%
\bibitem [{\citenamefont {Read}\ and\ \citenamefont
  {Green}()}]{read_paired_2000}%
  \BibitemOpen
  \bibfield  {author} {\bibinfo {author} {\bibfnamefont {N.}~\bibnamefont
  {Read}}\ and\ \bibinfo {author} {\bibfnamefont {D.}~\bibnamefont {Green}},\
  }\bibfield  {title} {\bibinfo {title} {Paired states of fermions in two
  dimensions with breaking of parity and time-reversal symmetries and the
  fractional quantum hall effect},\ }\href
  {https://doi.org/10.1103/PhysRevB.61.10267} {\bibfield  {journal} {\bibinfo
  {journal} {Physical Review B}\ }\textbf {\bibinfo {volume} {61}},\ \bibinfo
  {pages} {10267}}\BibitemShut {NoStop}%
\bibitem [{\citenamefont {Kitaev}()}]{kitaev_unpaired_2001}%
  \BibitemOpen
  \bibfield  {author} {\bibinfo {author} {\bibfnamefont {A.~Y.}\ \bibnamefont
  {Kitaev}},\ }\bibfield  {title} {\bibinfo {title} {Unpaired majorana fermions
  in quantum wires},\ }\href {https://doi.org/10.1070/1063-7869/44/10S/S29}
  {\bibfield  {journal} {\bibinfo  {journal} {Physics-Uspekhi}\ }\textbf
  {\bibinfo {volume} {44}},\ \bibinfo {pages} {131}}\BibitemShut {NoStop}%
\bibitem [{\citenamefont {Ivanov}()}]{ivanov_non-abelian_2001}%
  \BibitemOpen
  \bibfield  {author} {\bibinfo {author} {\bibfnamefont {D.~A.}\ \bibnamefont
  {Ivanov}},\ }\bibfield  {title} {\bibinfo {title} {Non-abelian statistics of
  half-quantum vortices in p -wave superconductors},\ }\href
  {https://doi.org/10.1103/PhysRevLett.86.268} {\bibfield  {journal} {\bibinfo
  {journal} {Physical Review Letters}\ }\textbf {\bibinfo {volume} {86}},\
  \bibinfo {pages} {268}}\BibitemShut {NoStop}%
\bibitem [{\citenamefont {Leggett}(2001)}]{Leggett_2001_SSB_SF}%
  \BibitemOpen
  \bibfield  {author} {\bibinfo {author} {\bibfnamefont {A.~J.}\ \bibnamefont
  {Leggett}},\ }\bibfield  {title} {\bibinfo {title} {Bose-einstein
  condensation in the alkali gases: Some fundamental concepts},\ }\href
  {https://doi.org/10.1103/RevModPhys.73.307} {\bibfield  {journal} {\bibinfo
  {journal} {Rev. Mod. Phys.}\ }\textbf {\bibinfo {volume} {73}},\ \bibinfo
  {pages} {307} (\bibinfo {year} {2001})}\BibitemShut {NoStop}%
\bibitem [{\citenamefont {Ortiz}\ \emph {et~al.}(2014)\citenamefont {Ortiz},
  \citenamefont {Dukelsky}, \citenamefont {Cobanera}, \citenamefont {Esebbag},\
  and\ \citenamefont {Beenakker}}]{ortizManyBody2014}%
  \BibitemOpen
  \bibfield  {author} {\bibinfo {author} {\bibfnamefont {G.}~\bibnamefont
  {Ortiz}}, \bibinfo {author} {\bibfnamefont {J.}~\bibnamefont {Dukelsky}},
  \bibinfo {author} {\bibfnamefont {E.}~\bibnamefont {Cobanera}}, \bibinfo
  {author} {\bibfnamefont {C.}~\bibnamefont {Esebbag}},\ and\ \bibinfo {author}
  {\bibfnamefont {C.}~\bibnamefont {Beenakker}},\ }\bibfield  {title} {\bibinfo
  {title} {Many-{{Body Characterization}} of {{Particle-Conserving Topological
  Superfluids}}},\ }\href {https://doi.org/10.1103/PhysRevLett.113.267002}
  {\bibfield  {journal} {\bibinfo  {journal} {Physical Review Letters}\
  }\textbf {\bibinfo {volume} {113}},\ \bibinfo {pages} {267002} (\bibinfo
  {year} {2014})}\BibitemShut {NoStop}%
\bibitem [{\citenamefont {Wang}\ \emph {et~al.}(2017)\citenamefont {Wang},
  \citenamefont {Xu}, \citenamefont {Pu},\ and\ \citenamefont
  {Hazzard}}]{wangNumberconserving2017}%
  \BibitemOpen
  \bibfield  {author} {\bibinfo {author} {\bibfnamefont {Z.}~\bibnamefont
  {Wang}}, \bibinfo {author} {\bibfnamefont {Y.}~\bibnamefont {Xu}}, \bibinfo
  {author} {\bibfnamefont {H.}~\bibnamefont {Pu}},\ and\ \bibinfo {author}
  {\bibfnamefont {K.~R.~A.}\ \bibnamefont {Hazzard}},\ }\bibfield  {title}
  {\bibinfo {title} {Number-conserving interacting fermion models with exact
  topological superconducting ground states},\ }\href
  {https://doi.org/10.1103/PhysRevB.96.115110} {\bibfield  {journal} {\bibinfo
  {journal} {Physical Review B}\ }\textbf {\bibinfo {volume} {96}},\ \bibinfo
  {pages} {115110} (\bibinfo {year} {2017})}\BibitemShut {NoStop}%
\bibitem [{\citenamefont {Knapp}\ \emph {et~al.}(2020)\citenamefont {Knapp},
  \citenamefont {V{\"a}yrynen},\ and\ \citenamefont
  {Lutchyn}}]{knappNumberconserving2020}%
  \BibitemOpen
  \bibfield  {author} {\bibinfo {author} {\bibfnamefont {C.}~\bibnamefont
  {Knapp}}, \bibinfo {author} {\bibfnamefont {J.~I.}\ \bibnamefont
  {V{\"a}yrynen}},\ and\ \bibinfo {author} {\bibfnamefont {R.~M.}\ \bibnamefont
  {Lutchyn}},\ }\bibfield  {title} {\bibinfo {title} {Number-conserving
  analysis of measurement-based braiding with {{Majorana}} zero modes},\ }\href
  {https://doi.org/10.1103/PhysRevB.101.125108} {\bibfield  {journal} {\bibinfo
   {journal} {Physical Review B}\ }\textbf {\bibinfo {volume} {101}},\ \bibinfo
  {pages} {125108} (\bibinfo {year} {2020})}\BibitemShut {NoStop}%
\bibitem [{\citenamefont {Lapa}\ and\ \citenamefont
  {Levin}(2020)}]{lapaRigorous2020}%
  \BibitemOpen
  \bibfield  {author} {\bibinfo {author} {\bibfnamefont {M.~F.}\ \bibnamefont
  {Lapa}}\ and\ \bibinfo {author} {\bibfnamefont {M.}~\bibnamefont {Levin}},\
  }\bibfield  {title} {\bibinfo {title} {Rigorous {{Results}} on {{Topological
  Superconductivity}} with {{Particle Number Conservation}}},\ }\href
  {https://doi.org/10.1103/PhysRevLett.124.257002} {\bibfield  {journal}
  {\bibinfo  {journal} {Physical Review Letters}\ }\textbf {\bibinfo {volume}
  {124}},\ \bibinfo {pages} {257002} (\bibinfo {year} {2020})}\BibitemShut
  {NoStop}%
\bibitem [{\citenamefont {Lin}\ and\ \citenamefont
  {Leggett}(2022)}]{lin2022some}%
  \BibitemOpen
  \bibfield  {author} {\bibinfo {author} {\bibfnamefont {Y.}~\bibnamefont
  {Lin}}\ and\ \bibinfo {author} {\bibfnamefont {A.}~\bibnamefont {Leggett}},\
  }\bibfield  {title} {\bibinfo {title} {Some questions concerning majorana
  fermions in 2d (p+ ip) fermi superfluids},\ }\href
  {https://link.springer.com/article/10.1007/s44214-022-00006-w} {\bibfield
  {journal} {\bibinfo  {journal} {Quantum Frontiers}\ }\textbf {\bibinfo
  {volume} {1}},\ \bibinfo {pages} {4} (\bibinfo {year} {2022})}\BibitemShut
  {NoStop}%
\bibitem [{\citenamefont {Richardson}\ and\ \citenamefont
  {Sherman}(1964)}]{richardsonExact1964}%
  \BibitemOpen
  \bibfield  {author} {\bibinfo {author} {\bibfnamefont {R.~W.}\ \bibnamefont
  {Richardson}}\ and\ \bibinfo {author} {\bibfnamefont {N.}~\bibnamefont
  {Sherman}},\ }\bibfield  {title} {\bibinfo {title} {Exact eigenstates of the
  pairing-force {{Hamiltonian}}},\ }\href
  {https://doi.org/10.1016/0029-5582(64)90687-X} {\bibfield  {journal}
  {\bibinfo  {journal} {Nuclear Physics}\ }\textbf {\bibinfo {volume} {52}},\
  \bibinfo {pages} {221} (\bibinfo {year} {1964})}\BibitemShut {NoStop}%
\bibitem [{\citenamefont {Bardeen}(1957)}]{bardeenGauge1957}%
  \BibitemOpen
  \bibfield  {author} {\bibinfo {author} {\bibfnamefont {J.}~\bibnamefont
  {Bardeen}},\ }\bibfield  {title} {\bibinfo {title} {Gauge invariance and the
  energy gap model of superconductivity},\ }\href
  {https://doi.org/10.1007/BF02856068} {\bibfield  {journal} {\bibinfo
  {journal} {Il Nuovo Cimento (1955-1965)}\ }\textbf {\bibinfo {volume} {5}},\
  \bibinfo {pages} {1766} (\bibinfo {year} {1957})}\BibitemShut {NoStop}%
\bibitem [{\citenamefont {Nambu}()}]{nambu_quasi-particles_1960}%
  \BibitemOpen
  \bibfield  {author} {\bibinfo {author} {\bibfnamefont {Y.}~\bibnamefont
  {Nambu}},\ }\bibfield  {title} {\bibinfo {title} {Quasi-particles and gauge
  invariance in the theory of superconductivity},\ }\href
  {https://doi.org/10.1103/PhysRev.117.648} {\bibfield  {journal} {\bibinfo
  {journal} {Physical Review}\ }\textbf {\bibinfo {volume} {117}},\ \bibinfo
  {pages} {648}}\BibitemShut {NoStop}%
\bibitem [{\citenamefont {Baym}\ and\ \citenamefont
  {Kadanoff}()}]{baym_conservation_1961}%
  \BibitemOpen
  \bibfield  {author} {\bibinfo {author} {\bibfnamefont {G.}~\bibnamefont
  {Baym}}\ and\ \bibinfo {author} {\bibfnamefont {L.~P.}\ \bibnamefont
  {Kadanoff}},\ }\bibfield  {title} {\bibinfo {title} {Conservation laws and
  correlation functions},\ }\href {https://doi.org/10.1103/PhysRev.124.287}
  {\bibfield  {journal} {\bibinfo  {journal} {Physical Review}\ }\textbf
  {\bibinfo {volume} {124}},\ \bibinfo {pages} {287}}\BibitemShut {NoStop}%
\bibitem [{\citenamefont {Bardeen}\ \emph {et~al.}(1957)\citenamefont
  {Bardeen}, \citenamefont {Cooper},\ and\ \citenamefont
  {Schrieffer}}]{bardeenTheory1957}%
  \BibitemOpen
  \bibfield  {author} {\bibinfo {author} {\bibfnamefont {J.}~\bibnamefont
  {Bardeen}}, \bibinfo {author} {\bibfnamefont {L.~N.}\ \bibnamefont
  {Cooper}},\ and\ \bibinfo {author} {\bibfnamefont {J.~R.}\ \bibnamefont
  {Schrieffer}},\ }\bibfield  {title} {\bibinfo {title} {Theory of
  {{Superconductivity}}},\ }\href {https://doi.org/10.1103/PhysRev.108.1175}
  {\bibfield  {journal} {\bibinfo  {journal} {Physical Review}\ }\textbf
  {\bibinfo {volume} {108}},\ \bibinfo {pages} {1175} (\bibinfo {year}
  {1957})}\BibitemShut {NoStop}%
\bibitem [{\citenamefont {Bogoljubov}()}]{bogoljubovnew1958}%
  \BibitemOpen
  \bibfield  {author} {\bibinfo {author} {\bibfnamefont {N.~N.}\ \bibnamefont
  {Bogoljubov}},\ }\bibfield  {title} {\bibinfo {title} {On a new method in the
  theory of superconductivity},\ }\href {https://doi.org/10.1007/BF02745585}
  {\bibfield  {journal} {\bibinfo  {journal} {Il Nuovo Cimento (1955-1965)}\
  }\textbf {\bibinfo {volume} {7}},\ \bibinfo {pages} {794}}\BibitemShut
  {NoStop}%
\bibitem [{\citenamefont {Richard}\ \emph {et~al.}(2015)\citenamefont
  {Richard}, \citenamefont {Qian},\ and\ \citenamefont
  {Ding}}]{Richard2015ARPESMO}%
  \BibitemOpen
  \bibfield  {author} {\bibinfo {author} {\bibfnamefont {P.}~\bibnamefont
  {Richard}}, \bibinfo {author} {\bibfnamefont {T.}~\bibnamefont {Qian}},\ and\
  \bibinfo {author} {\bibfnamefont {H.}~\bibnamefont {Ding}},\ }\bibfield
  {title} {\bibinfo {title} {Arpes measurements of the superconducting gap of
  fe-based superconductors and their implications to the pairing mechanism},\
  }\href {https://api.semanticscholar.org/CorpusID:32145016} {\bibfield
  {journal} {\bibinfo  {journal} {Journal of Physics: Condensed Matter}\
  }\textbf {\bibinfo {volume} {27}} (\bibinfo {year} {2015})}\BibitemShut
  {NoStop}%
\bibitem [{\citenamefont {Balatsky}\ \emph {et~al.}(2009)\citenamefont
  {Balatsky}, \citenamefont {Lee},\ and\ \citenamefont
  {Shen}}]{PhysRevB.79.020505}%
  \BibitemOpen
  \bibfield  {author} {\bibinfo {author} {\bibfnamefont {A.~V.}\ \bibnamefont
  {Balatsky}}, \bibinfo {author} {\bibfnamefont {W.~S.}\ \bibnamefont {Lee}},\
  and\ \bibinfo {author} {\bibfnamefont {Z.~X.}\ \bibnamefont {Shen}},\
  }\bibfield  {title} {\bibinfo {title} {Bogoliubov angle, particle-hole
  mixture, and angle-resolved photoemission spectroscopy in superconductors},\
  }\href {https://doi.org/10.1103/PhysRevB.79.020505} {\bibfield  {journal}
  {\bibinfo  {journal} {Phys. Rev. B}\ }\textbf {\bibinfo {volume} {79}},\
  \bibinfo {pages} {020505} (\bibinfo {year} {2009})}\BibitemShut {NoStop}%
\bibitem [{\citenamefont {Damascelli}\ \emph {et~al.}(2003)\citenamefont
  {Damascelli}, \citenamefont {Hussain},\ and\ \citenamefont
  {Shen}}]{RevModPhys.75.473}%
  \BibitemOpen
  \bibfield  {author} {\bibinfo {author} {\bibfnamefont {A.}~\bibnamefont
  {Damascelli}}, \bibinfo {author} {\bibfnamefont {Z.}~\bibnamefont
  {Hussain}},\ and\ \bibinfo {author} {\bibfnamefont {Z.-X.}\ \bibnamefont
  {Shen}},\ }\bibfield  {title} {\bibinfo {title} {Angle-resolved photoemission
  studies of the cuprate superconductors},\ }\href
  {https://doi.org/10.1103/RevModPhys.75.473} {\bibfield  {journal} {\bibinfo
  {journal} {Rev. Mod. Phys.}\ }\textbf {\bibinfo {volume} {75}},\ \bibinfo
  {pages} {473} (\bibinfo {year} {2003})}\BibitemShut {NoStop}%
\bibitem [{\citenamefont {Fischer}\ \emph {et~al.}(2007)\citenamefont
  {Fischer}, \citenamefont {Kugler}, \citenamefont {Maggio-Aprile},
  \citenamefont {Berthod},\ and\ \citenamefont {Renner}}]{RevModPhys.79.353}%
  \BibitemOpen
  \bibfield  {author} {\bibinfo {author} {\bibfnamefont {O.}~\bibnamefont
  {Fischer}}, \bibinfo {author} {\bibfnamefont {M.}~\bibnamefont {Kugler}},
  \bibinfo {author} {\bibfnamefont {I.}~\bibnamefont {Maggio-Aprile}}, \bibinfo
  {author} {\bibfnamefont {C.}~\bibnamefont {Berthod}},\ and\ \bibinfo {author}
  {\bibfnamefont {C.}~\bibnamefont {Renner}},\ }\bibfield  {title} {\bibinfo
  {title} {Scanning tunneling spectroscopy of high-temperature
  superconductors},\ }\href {https://doi.org/10.1103/RevModPhys.79.353}
  {\bibfield  {journal} {\bibinfo  {journal} {Rev. Mod. Phys.}\ }\textbf
  {\bibinfo {volume} {79}},\ \bibinfo {pages} {353} (\bibinfo {year}
  {2007})}\BibitemShut {NoStop}%
\bibitem [{\citenamefont {Lee}\ \emph {et~al.}(2009)\citenamefont {Lee},
  \citenamefont {Fujita}, \citenamefont {Schmidt}, \citenamefont {Kim},
  \citenamefont {Eisaki}, \citenamefont {Uchida},\ and\ \citenamefont
  {Davis}}]{leespectroscopic2009}%
  \BibitemOpen
  \bibfield  {author} {\bibinfo {author} {\bibfnamefont {J.}~\bibnamefont
  {Lee}}, \bibinfo {author} {\bibfnamefont {K.}~\bibnamefont {Fujita}},
  \bibinfo {author} {\bibfnamefont {A.}~\bibnamefont {Schmidt}}, \bibinfo
  {author} {\bibfnamefont {C.~K.}\ \bibnamefont {Kim}}, \bibinfo {author}
  {\bibfnamefont {H.}~\bibnamefont {Eisaki}}, \bibinfo {author} {\bibfnamefont
  {S.}~\bibnamefont {Uchida}},\ and\ \bibinfo {author} {\bibfnamefont
  {J.}~\bibnamefont {Davis}},\ }\bibfield  {title} {\bibinfo {title}
  {Spectroscopic fingerprint of phase-incoherent superconductivity in the
  underdoped bi2sr2cacu2o8+ $\delta$},\ }\href@noop {} {\bibfield  {journal}
  {\bibinfo  {journal} {Science}\ }\textbf {\bibinfo {volume} {325}},\ \bibinfo
  {pages} {1099} (\bibinfo {year} {2009})}\BibitemShut {NoStop}%
\bibitem [{\citenamefont {Deutscher}(2005)}]{RevModPhys.77.109}%
  \BibitemOpen
  \bibfield  {author} {\bibinfo {author} {\bibfnamefont {G.}~\bibnamefont
  {Deutscher}},\ }\bibfield  {title} {\bibinfo {title} {Andreev--saint-james
  reflections: A probe of cuprate superconductors},\ }\href
  {https://doi.org/10.1103/RevModPhys.77.109} {\bibfield  {journal} {\bibinfo
  {journal} {Rev. Mod. Phys.}\ }\textbf {\bibinfo {volume} {77}},\ \bibinfo
  {pages} {109} (\bibinfo {year} {2005})}\BibitemShut {NoStop}%
\bibitem [{sup()}]{supplementary}%
  \BibitemOpen
  \href@noop {} {}\bibinfo {note} {See supplementary Material for the
  disscussion 1. One-body Green's function represented by eigen-particles, 2.
  Particle conservation in systems with spontaneously broken symmetry, 3.
  Proper and improper Bogoliubov quasi-particles}\BibitemShut {NoStop}%
\bibitem [{\citenamefont {Anderson}(1972)}]{MoreAnderson1972}%
  \BibitemOpen
  \bibfield  {author} {\bibinfo {author} {\bibfnamefont {P.~W.}\ \bibnamefont
  {Anderson}},\ }\bibfield  {title} {\bibinfo {title} {More is different},\
  }\href {https://doi.org/10.1126/science.177.4047.393} {\bibfield  {journal}
  {\bibinfo  {journal} {Science}\ }\textbf {\bibinfo {volume} {177}},\ \bibinfo
  {pages} {393} (\bibinfo {year} {1972})}\BibitemShut {NoStop}%
\bibitem [{\citenamefont {Mattuck}\ and\ \citenamefont
  {Chang}(1976)}]{Mattuck1976AGT}%
  \BibitemOpen
  \bibfield  {author} {\bibinfo {author} {\bibfnamefont {R.~D.}\ \bibnamefont
  {Mattuck}}\ and\ \bibinfo {author} {\bibfnamefont {H.~H.~C.}\ \bibnamefont
  {Chang}},\ }\href@noop {} {\emph {\bibinfo {title} {A guide to Feynman
  diagrams in the many-body problem}}}\ (\bibinfo  {publisher} {Courier
  Corporation},\ \bibinfo {year} {1976})\BibitemShut {NoStop}%
\bibitem [{\citenamefont {Anderson}(1997)}]{andersonconcepts1997}%
  \BibitemOpen
  \bibfield  {author} {\bibinfo {author} {\bibfnamefont {P.~W.}\ \bibnamefont
  {Anderson}},\ }\href@noop {} {\emph {\bibinfo {title} {Concepts in solids:
  lectures on the theory of solids}}},\ Vol.~\bibinfo {volume} {58}\ (\bibinfo
  {publisher} {World Scientific},\ \bibinfo {year} {1997})\BibitemShut
  {NoStop}%
\bibitem [{\citenamefont {White}(2002)}]{whiteNumerical2002}%
  \BibitemOpen
  \bibfield  {author} {\bibinfo {author} {\bibfnamefont {S.~R.}\ \bibnamefont
  {White}},\ }\bibfield  {title} {\bibinfo {title} {Numerical canonical
  transformation approach to quantum many-body problems},\ }\href
  {https://doi.org/10.1063/1.1508370} {\bibfield  {journal} {\bibinfo
  {journal} {The Journal of Chemical Physics}\ }\textbf {\bibinfo {volume}
  {117}},\ \bibinfo {pages} {7472} (\bibinfo {year} {2002})}\BibitemShut
  {NoStop}%
\bibitem [{\citenamefont {Kanno}()}]{kanno_method_1969}%
  \BibitemOpen
  \bibfield  {author} {\bibinfo {author} {\bibfnamefont {S.}~\bibnamefont
  {Kanno}},\ }\bibfield  {title} {\bibinfo {title} {A method of a quasi-linear
  canonical transformation for the many-body problem. i},\ }\href
  {https://doi.org/10.1143/PTP.41.966} {\bibfield  {journal} {\bibinfo
  {journal} {Progress of Theoretical Physics}\ }\textbf {\bibinfo {volume}
  {41}},\ \bibinfo {pages} {966}}\BibitemShut {NoStop}%
\bibitem [{\citenamefont {Hegg}\ \emph {et~al.}(2024)\citenamefont {Hegg},
  \citenamefont {Jiang}, \citenamefont {Wang}, \citenamefont {Hou},
  \citenamefont {Zeng}, \citenamefont {Yildirim},\ and\ \citenamefont
  {Ku}}]{heggUniversal2024}%
  \BibitemOpen
  \bibfield  {author} {\bibinfo {author} {\bibfnamefont {A.}~\bibnamefont
  {Hegg}}, \bibinfo {author} {\bibfnamefont {R.}~\bibnamefont {Jiang}},
  \bibinfo {author} {\bibfnamefont {J.}~\bibnamefont {Wang}}, \bibinfo {author}
  {\bibfnamefont {J.}~\bibnamefont {Hou}}, \bibinfo {author} {\bibfnamefont
  {T.}~\bibnamefont {Zeng}}, \bibinfo {author} {\bibfnamefont {Y.}~\bibnamefont
  {Yildirim}},\ and\ \bibinfo {author} {\bibfnamefont {W.}~\bibnamefont {Ku}},\
  }\href {https://doi.org/10.48550/arXiv.2402.08730} {\bibinfo {title}
  {Universal low-temperature fluctuation of unconventional superconductors
  revealed: '{{Smoking}} gun' leaves proper bosonic superfluidity the last
  theory standing}} (\bibinfo {year} {2024}),\ \Eprint
  {https://arxiv.org/abs/2402.08730} {arXiv:2402.08730 [cond-mat]} \BibitemShut
  {NoStop}%
\bibitem [{\citenamefont {Schrieffer}\ and\ \citenamefont
  {Wolff}(1966)}]{SW_transform}%
  \BibitemOpen
  \bibfield  {author} {\bibinfo {author} {\bibfnamefont {J.~R.}\ \bibnamefont
  {Schrieffer}}\ and\ \bibinfo {author} {\bibfnamefont {P.~A.}\ \bibnamefont
  {Wolff}},\ }\bibfield  {title} {\bibinfo {title} {Relation between the
  anderson and kondo hamiltonians},\ }\href
  {https://doi.org/10.1103/PhysRev.149.491} {\bibfield  {journal} {\bibinfo
  {journal} {Phys. Rev.}\ }\textbf {\bibinfo {volume} {149}},\ \bibinfo {pages}
  {491} (\bibinfo {year} {1966})}\BibitemShut {NoStop}%
\bibitem [{\citenamefont {Kivelson}\ and\ \citenamefont
  {Rokhsar}(1990)}]{KivelsonBogoliubov1990}%
  \BibitemOpen
  \bibfield  {author} {\bibinfo {author} {\bibfnamefont {S.~A.}\ \bibnamefont
  {Kivelson}}\ and\ \bibinfo {author} {\bibfnamefont {D.~S.}\ \bibnamefont
  {Rokhsar}},\ }\bibfield  {title} {\bibinfo {title} {Bogoliubov
  quasiparticles, spinons, and spin-charge decoupling in superconductors},\
  }\href {https://doi.org/10.1103/PhysRevB.41.11693} {\bibfield  {journal}
  {\bibinfo  {journal} {Phys. Rev. B}\ }\textbf {\bibinfo {volume} {41}},\
  \bibinfo {pages} {11693} (\bibinfo {year} {1990})}\BibitemShut {NoStop}%
\bibitem [{\citenamefont {Hansson}\ \emph {et~al.}(2004)\citenamefont
  {Hansson}, \citenamefont {Oganesyan},\ and\ \citenamefont
  {Sondhi}}]{hanssonsuperconductors2004}%
  \BibitemOpen
  \bibfield  {author} {\bibinfo {author} {\bibfnamefont {T.}~\bibnamefont
  {Hansson}}, \bibinfo {author} {\bibfnamefont {V.}~\bibnamefont {Oganesyan}},\
  and\ \bibinfo {author} {\bibfnamefont {S.~L.}\ \bibnamefont {Sondhi}},\
  }\bibfield  {title} {\bibinfo {title} {Superconductors are topologically
  ordered},\ }\href@noop {} {\bibfield  {journal} {\bibinfo  {journal} {Annals
  of Physics}\ }\textbf {\bibinfo {volume} {313}},\ \bibinfo {pages} {497}
  (\bibinfo {year} {2004})}\BibitemShut {NoStop}%
\bibitem [{\citenamefont {Giamarchi}(2003)}]{giamarchi2003quantum}%
  \BibitemOpen
  \bibfield  {author} {\bibinfo {author} {\bibfnamefont {T.}~\bibnamefont
  {Giamarchi}},\ }\href@noop {} {\emph {\bibinfo {title} {Quantum physics in
  one dimension}}},\ Vol.\ \bibinfo {volume} {121}\ (\bibinfo  {publisher}
  {Clarendon press},\ \bibinfo {year} {2003})\BibitemShut {NoStop}%
\bibitem [{\citenamefont {Sau}\ \emph {et~al.}(2011)\citenamefont {Sau},
  \citenamefont {Halperin}, \citenamefont {Flensberg},\ and\ \citenamefont
  {Das~Sarma}}]{NumberSau2011}%
  \BibitemOpen
  \bibfield  {author} {\bibinfo {author} {\bibfnamefont {J.~D.}\ \bibnamefont
  {Sau}}, \bibinfo {author} {\bibfnamefont {B.~I.}\ \bibnamefont {Halperin}},
  \bibinfo {author} {\bibfnamefont {K.}~\bibnamefont {Flensberg}},\ and\
  \bibinfo {author} {\bibfnamefont {S.}~\bibnamefont {Das~Sarma}},\ }\bibfield
  {title} {\bibinfo {title} {Number conserving theory for topologically
  protected degeneracy in one-dimensional fermions},\ }\href
  {https://doi.org/10.1103/PhysRevB.84.144509} {\bibfield  {journal} {\bibinfo
  {journal} {Phys. Rev. B}\ }\textbf {\bibinfo {volume} {84}},\ \bibinfo
  {pages} {144509} (\bibinfo {year} {2011})}\BibitemShut {NoStop}%
\bibitem [{\citenamefont {White}\ \emph {et~al.}(1989)\citenamefont {White},
  \citenamefont {Scalapino}, \citenamefont {Sugar}, \citenamefont {Bickers},\
  and\ \citenamefont {Scalettar}}]{AttractiveWhite1989}%
  \BibitemOpen
  \bibfield  {author} {\bibinfo {author} {\bibfnamefont {S.~R.}\ \bibnamefont
  {White}}, \bibinfo {author} {\bibfnamefont {D.~J.}\ \bibnamefont
  {Scalapino}}, \bibinfo {author} {\bibfnamefont {R.~L.}\ \bibnamefont
  {Sugar}}, \bibinfo {author} {\bibfnamefont {N.~E.}\ \bibnamefont {Bickers}},\
  and\ \bibinfo {author} {\bibfnamefont {R.~T.}\ \bibnamefont {Scalettar}},\
  }\bibfield  {title} {\bibinfo {title} {Attractive and repulsive pairing
  interaction vertices for the two-dimensional hubbard model},\ }\href
  {https://doi.org/10.1103/PhysRevB.39.839} {\bibfield  {journal} {\bibinfo
  {journal} {Phys. Rev. B}\ }\textbf {\bibinfo {volume} {39}},\ \bibinfo
  {pages} {839(R)} (\bibinfo {year} {1989})}\BibitemShut {NoStop}%
\bibitem [{\citenamefont {Luk}\ and\ \citenamefont
  {Cox}(1990)}]{SuperconductivityLuk1990}%
  \BibitemOpen
  \bibfield  {author} {\bibinfo {author} {\bibfnamefont {K.-H.}\ \bibnamefont
  {Luk}}\ and\ \bibinfo {author} {\bibfnamefont {D.~L.}\ \bibnamefont {Cox}},\
  }\bibfield  {title} {\bibinfo {title} {Superconductivity, hole motion, and
  spin-charge correlations in the t-j model in two dimensions},\ }\href
  {https://doi.org/10.1103/PhysRevB.41.4456} {\bibfield  {journal} {\bibinfo
  {journal} {Phys. Rev. B}\ }\textbf {\bibinfo {volume} {41}},\ \bibinfo
  {pages} {4456} (\bibinfo {year} {1990})}\BibitemShut {NoStop}%
\bibitem [{\citenamefont {White}\ and\ \citenamefont
  {Scalapino}(1997)}]{HoleWhite1997}%
  \BibitemOpen
  \bibfield  {author} {\bibinfo {author} {\bibfnamefont {S.~R.}\ \bibnamefont
  {White}}\ and\ \bibinfo {author} {\bibfnamefont {D.~J.}\ \bibnamefont
  {Scalapino}},\ }\bibfield  {title} {\bibinfo {title} {Hole and pair
  structures in the t-j model},\ }\href
  {https://doi.org/10.1103/PhysRevB.55.6504} {\bibfield  {journal} {\bibinfo
  {journal} {Phys. Rev. B}\ }\textbf {\bibinfo {volume} {55}},\ \bibinfo
  {pages} {6504} (\bibinfo {year} {1997})}\BibitemShut {NoStop}%
\bibitem [{\citenamefont {Moore}\ and\ \citenamefont
  {Read}(1991)}]{mooreNonabelions1991}%
  \BibitemOpen
  \bibfield  {author} {\bibinfo {author} {\bibfnamefont {G.}~\bibnamefont
  {Moore}}\ and\ \bibinfo {author} {\bibfnamefont {N.}~\bibnamefont {Read}},\
  }\bibfield  {title} {\bibinfo {title} {Nonabelions in the fractional quantum
  hall effect},\ }\href {https://doi.org/10.1016/0550-3213(91)90407-O}
  {\bibfield  {journal} {\bibinfo  {journal} {Nuclear Physics B}\ }\textbf
  {\bibinfo {volume} {360}},\ \bibinfo {pages} {362} (\bibinfo {year}
  {1991})}\BibitemShut {NoStop}%
\bibitem [{\citenamefont {Deutsch}(1991)}]{DeutschETH1991}%
  \BibitemOpen
  \bibfield  {author} {\bibinfo {author} {\bibfnamefont {J.~M.}\ \bibnamefont
  {Deutsch}},\ }\bibfield  {title} {\bibinfo {title} {Quantum statistical
  mechanics in a closed system},\ }\href
  {https://doi.org/10.1103/PhysRevA.43.2046} {\bibfield  {journal} {\bibinfo
  {journal} {Phys. Rev. A}\ }\textbf {\bibinfo {volume} {43}},\ \bibinfo
  {pages} {2046} (\bibinfo {year} {1991})}\BibitemShut {NoStop}%
\bibitem [{\citenamefont {Rigol}\ and\ \citenamefont
  {Srednicki}(2012)}]{SrednickiETH2012}%
  \BibitemOpen
  \bibfield  {author} {\bibinfo {author} {\bibfnamefont {M.}~\bibnamefont
  {Rigol}}\ and\ \bibinfo {author} {\bibfnamefont {M.}~\bibnamefont
  {Srednicki}},\ }\bibfield  {title} {\bibinfo {title} {Alternatives to
  eigenstate thermalization},\ }\href
  {https://doi.org/10.1103/PhysRevLett.108.110601} {\bibfield  {journal}
  {\bibinfo  {journal} {Phys. Rev. Lett.}\ }\textbf {\bibinfo {volume} {108}},\
  \bibinfo {pages} {110601} (\bibinfo {year} {2012})}\BibitemShut {NoStop}%
\bibitem [{\citenamefont {Holstein}\ and\ \citenamefont
  {Primakoff}(1940)}]{holsteinfield1940}%
  \BibitemOpen
  \bibfield  {author} {\bibinfo {author} {\bibfnamefont {T.}~\bibnamefont
  {Holstein}}\ and\ \bibinfo {author} {\bibfnamefont {H.}~\bibnamefont
  {Primakoff}},\ }\bibfield  {title} {\bibinfo {title} {Field dependence of the
  intrinsic domain magnetization of a ferromagnet},\ }\href@noop {} {\bibfield
  {journal} {\bibinfo  {journal} {Physical Review}\ }\textbf {\bibinfo {volume}
  {58}},\ \bibinfo {pages} {1098} (\bibinfo {year} {1940})}\BibitemShut
  {NoStop}%
\bibitem [{\citenamefont {Leggett}(2006)}]{leggett2006quantum}%
  \BibitemOpen
  \bibfield  {author} {\bibinfo {author} {\bibfnamefont {A.~J.}\ \bibnamefont
  {Leggett}},\ }\href@noop {} {\emph {\bibinfo {title} {Quantum liquids: Bose
  condensation and Cooper pairing in condensed-matter systems}}}\ (\bibinfo
  {publisher} {Oxford university press},\ \bibinfo {year} {2006})\BibitemShut
  {NoStop}%
\bibitem [{\citenamefont {Schrieffer}(2018)}]{schrieffertheory2018}%
  \BibitemOpen
  \bibfield  {author} {\bibinfo {author} {\bibfnamefont {J.~R.}\ \bibnamefont
  {Schrieffer}},\ }\href@noop {} {\emph {\bibinfo {title} {Theory of
  superconductivity}}}\ (\bibinfo  {publisher} {CRC press},\ \bibinfo {year}
  {2018})\ Chap.~\bibinfo {chapter} {2}\BibitemShut {NoStop}%
\bibitem [{\citenamefont {Blonder}\ \emph {et~al.}(1982)\citenamefont
  {Blonder}, \citenamefont {Tinkham},\ and\ \citenamefont
  {Klapwijk}}]{blonderTransition1982}%
  \BibitemOpen
  \bibfield  {author} {\bibinfo {author} {\bibfnamefont {G.~E.}\ \bibnamefont
  {Blonder}}, \bibinfo {author} {\bibfnamefont {M.}~\bibnamefont {Tinkham}},\
  and\ \bibinfo {author} {\bibfnamefont {T.~M.}\ \bibnamefont {Klapwijk}},\
  }\bibfield  {title} {\bibinfo {title} {Transition from metallic to tunneling
  regimes in superconducting microconstrictions: {{Excess}} current, charge
  imbalance, and supercurrent conversion},\ }\href
  {https://doi.org/10.1103/PhysRevB.25.4515} {\bibfield  {journal} {\bibinfo
  {journal} {Physical Review B}\ }\textbf {\bibinfo {volume} {25}},\ \bibinfo
  {pages} {4515} (\bibinfo {year} {1982})}\BibitemShut {NoStop}%
\bibitem [{\citenamefont {Tinkham}(2004)}]{tinkhamintron2004}%
  \BibitemOpen
  \bibfield  {author} {\bibinfo {author} {\bibfnamefont {M.}~\bibnamefont
  {Tinkham}},\ }\href@noop {} {\emph {\bibinfo {title} {Introduction to
  superconductivity}}}\ (\bibinfo  {publisher} {Courier Corporation},\ \bibinfo
  {year} {2004})\BibitemShut {NoStop}%
\end{thebibliography}%

\onecolumngrid
\clearpage
\appendix
\section*{Supplementary Materials}

\setcounter{equation}{0}
\setcounter{figure}{0}
\setcounter{table}{0}
\setcounter{page}{1}
\makeatletter
\renewcommand{\theequation}{S\arabic{equation}}
\renewcommand{\thefigure}{S\arabic{figure}}
\renewcommand{\thetable}{S\arabic{table}}
\renewcommand{\bibnumfmt}[1]{[S#1]}
\renewcommand{\citenumfont}[1]{S#1}

\section{1. One-body Green's function represented by eigen-particles}
In this section we expand the analytical structure of eigen-particle representation of bare particles' one-body Green's function, which is typically accessible via angular-resolve photo-emission experiments.
From this, one can see that typical one-body quasi-particle peaks acquire their energy through the eign-particle energies, while integration over their fluctuations produces the observed continuum and peak broadening.

Given the canonical transformation, $\mathcal{U}$, of particles, $\tilde{c}^\dagger_l \equiv \mathcal{U}^\dagger c^\dagger_l \mathcal{U}$, that diagonalizes the bare Hamiltonian, $H$,
\[
\tilde{H}[\{\tilde{c}^\dagger_l\}]\equiv H[\{c^\dagger_l\}]=\mathcal{U}H[\{\tilde{c}^\dagger_l\}]\mathcal{U}^{\dagger}=\sum_{l_{1}}\epsilon_{l_{1}}\tilde{c}_{l_{1}}^{\dagger}\tilde{c}_{l_{1}}+\frac{1}{2!}\sum_{l_{1}l_{2}}\epsilon_{l_{1}l_{2}}\tilde{c}_{l_{1}}^{\dagger}\tilde{c}_{l_{2}}^{\dagger}\tilde{c}_{l_{2}}\tilde{c}_{l_{1}}+\cdots,
\]
the bare particles can be represented by the eigen-particles through the inverse transform,
\begin{align*}
c^\dagger_{l} &= \mathcal{U}
\tilde{c}^\dagger_{l}\mathcal{U}^\dagger=\tilde{c}_{l}^{\dagger}+\frac{1}{2!}\sum_{l_1 l_2 l_1'} A_{l_1 l_2; l l_1'}\tilde{c}_{l_1}^{\dagger}\tilde{c}_{l_2}^{\dagger}  \tilde{c}_{l_1'}+\cdots,
\end{align*}
where coefficients $A_{l_1 l_2; l l_1'}$ satisfies the anti-permutation, $A_{l_1 l_2; l l_1'}=-A_{l_2 l_1; l l_1'}=-A_{l_1 l_2; l_1' l}$, as required by the fermionic canonical condition, $\{\tilde{c}_l,\tilde{c}_{l^\prime}\}=\{c_l,c_{l^\prime}\}=0$~\cite{kanno_method_1969}.
Given all many-body eigen-states as product states of eigen-particles, the time-ordered one-body Green's function \textit{operator} for bare particle,
\begin{align*}
G_{l}(t) & \equiv-i\mathcal{T}[c_{l}(t)c_{l}^{\dagger}(0)]\\
 & =-i[\theta(t)c_{l}(t)c_{l}^{\dagger}(0)-\theta(-t)c_{l}^{\dagger}(0)c_{l}(t)],
\end{align*}
describes the probability of adding an electron (first term) or removing one (second term, equivalent to adding a hole) can be \textit{directly} accessed in eigen-particle representation, analogous to (but more informative than) the standard Lehmann representation.

As an illustration, the component in the first term reads,
\begin{align*}
    c_{l}(t)c_{l}^{\dagger}(0)	=&(\tilde{c}_{l}+\frac{1}{2!}\sum_{m_{1}m_{2}m_{1}'}A_{m_{1}m_{2};lm_{1}'}^{*}\tilde{c}_{m_{1}'}^{\dagger}\tilde{c}_{m_{2}}\tilde{c}_{m_{1}}+\cdots)_{t}(\tilde{c}_{l}^{\dagger}+\frac{1}{2!}\sum_{l_{1}l_{2}l_{1}'}A_{l_{1}l_{2};ll_{1}'}\tilde{c}_{l_{1}}^{\dagger}\tilde{c}_{l_{2}}^{\dagger}\tilde{c}_{l_{1}'}+\cdots)_{0}\\
	=&(e^{-iE_{l}t}\tilde{c}_{l}+\frac{1}{2!}\sum_{m_{1}m_{2}m_{1}'}e^{-iE_{m_{1}m_{2};m_{1'}}t}A_{m_{1}m_{2};lm_{1}'}^{*}\tilde{c}_{m_{1}'}^{\dagger}\tilde{c}_{m_{2}}\tilde{c}_{m_{1}})(\tilde{c}_{l}^{\dagger}+\frac{1}{2!}\sum_{l_{1}l_{2}l_{1}'}A_{l_{1}l_{2};ll_{1}'}\tilde{c}_{l_{1}}^{\dagger}\tilde{c}_{l_{2}}^{\dagger}\tilde{c}_{l_{1}'}+\cdots)\\
	=&e^{-iE_{l}t}\tilde{c}_{l}\tilde{c}_{l}^{\dagger}+\frac{1}{2!}\sum_{l_{1}l_{2}l_{1}'}e^{-iE_{l}t}A_{l_{1}l_{2};ll_{1}'}\tilde{c}_{l}\tilde{c}_{l_{1}}^{\dagger}\tilde{c}_{l_{2}}^{\dagger}\tilde{c}_{l_{1}'}+\frac{1}{2!}\sum_{m_{1}m_{2}m_{1}'}e^{-iE_{m_{1}m_{2};m_{1'}}t}A_{m_{1}m_{2};lm_{1}'}^{*}\tilde{c}_{m_1'}^{\dagger}\tilde{c}_{m_{2}}\tilde{c}_{m_{1}}\tilde{c}_{l}^{\dagger}\\
	+&\frac{1}{(2!)^{2}}\sum_{m_{1}m_{2}m_{1}'}\sum_{l_{1}l_{2}l_{1}'}e^{-iE_{m_{1}m_{2};m_{1}'}t}A_{m_{1}m_{2};lm_{1}'}^{*}A_{l_{1}l_{2};ll_{1}'}\tilde{c}_{m_{1}'}^{\dagger}\tilde{c}_{m_{2}}\tilde{c}_{m_{1}}\tilde{c}_{l_{1}}^{\dagger}\tilde{c}_{l_{2}}^{\dagger}\tilde{c}_{l_{1}'}+\cdots,
\end{align*}
where
\begin{align}
&[H,\tilde{c}_{l}^{\dagger}]	\equiv\tilde{c}_{l}^{\dagger}E_{l};\\
&[H,\tilde{c}_{m_{1}}^{\dagger}\tilde{c}_{m_{2}}^{\dagger}\tilde{c}_{m_1'}]	\equiv\tilde{c}_{m_{1}}^{\dagger}\tilde{c}_{m_{2}}^{\dagger}\tilde{c}_{m_1'}E_{m_{1}m_{2};m_1'}.
\label{qp_energy}
\end{align}
are diagonal Hermitian energy \textit{operators} for $\tilde{c}^\dagger_l$ and the composite objects, $\tilde{c}^\dagger_{m_1} \tilde{c}^\dagger_{m_2} \tilde{c}_{m^\prime_1}$.
In turn, the corresponding contribution to the frequency-dependent one-body Green's function operator, $G_l(\omega)\equiv\int dt G_l(t) e^{i\omega t}$, for \textit{any} eigen-state is directly given by the diagonal operator,
\begin{align*}
		&(\omega-E_{l}+i0^{+})^{-1}\tilde{c}_{l}\tilde{c}_{l}^{\dagger}+\sum_{l_{1}}(\omega-E_{l}+i0^{+})^{-1}A_{ll_{1};ll_{1}}\tilde{c}_{l}\tilde{c}_{l}^{\dagger}\tilde{c}_{l_{1}}^{\dagger}\tilde{c}_{l_{1}}+\sum_{m_{1}}(\omega-E_{m_{1}l;m_{1}}+i0^{+})^{-1}A_{lm_{1};lm_{1}}^{*}\tilde{c}_{m_{1}}^{\dagger}\tilde{c}_{m_{1}}\tilde{c}_{l}\tilde{c}_{l}^{\dagger}\\
	&+\sum_{m_{1}l_{1}l_{2}}(\omega-E_{m_{1}l_{2};m_{1}}+i0^{+})^{-1}A_{lm_{1};m_{1}l_{2}}^{*}A_{l_{1}l_{2};ll_{1}}\tilde{c}_{m_{1}}^{\dagger}\tilde{c}_{l_{2}}\tilde{c}_{m_{1}}\tilde{c}_{l_{2}}^{\dagger}\tilde{c}_{l_{1}}^{\dagger}\tilde{c}_{l_{1}}\\
    &+\frac{1}{2!}\sum_{m_{1}m_{2}m_{1}'}(\omega-E_{m_{1}m_{2};m_{1'}}+i0^{+})^{-1}A_{m_{1}m_{2};lm_{1}'}^{*}A_{m_{1}m_{2};lm_{1}'}\tilde{c}_{m_{1}'}^{\dagger}\tilde{c}_{m_{2}}\tilde{c}_{m_{1}}\tilde{c}_{m_{1}}^{\dagger}\tilde{c}_{m_{2}}^{\dagger}\tilde{c}_{m_{1}'}+\cdots.
\end{align*}
Note $E_{m_1 l_2; m_1} = E_{l_2}$ from its definition in Eq.\ref{qp_energy}. This means that physically, the weight of quasi-particle peaks can be depleted through higher-order processes.

The same derivation applies to the hole-adding contribution as well, resulting in full $G_{l}(\omega)$ as
\begin{align*}
G_{l}(\omega)=&	(\omega-E_{l}+i0^{+})^{-1}\tilde{c}_{l}\tilde{c}_{l}^{\dagger}\left[1+\sum_{l_{1}}A_{ll_{1};ll_{1}}\tilde{c}_{l_{1}}^{\dagger}\tilde{c}_{l_{1}}+\sum_{m_{1}}A_{lm_{1};lm_{1}}^{*}\tilde{c}_{m_{1}}^{\dagger}\tilde{c}_{m_{1}}+\sum_{m_{1}l_{1}}A_{lm_{1};m_{1}l}^{*}A_{l_{1}l;ll_{1}}\tilde{c}_{m_{1}}^{\dagger}\tilde{c}_{m_{1}}\tilde{c}_{l_{1}}^{\dagger}\tilde{c}_{l_{1}}+\cdots\right]\\
	+&\sum_{m_{1}l_{1}}\sum_{l_{2}\neq l}(\omega-E_{l_{2}}+i0^{+})^{-1}A_{lm_{1};m_{1}l_{2}}^{*}A_{l_{1}l_{2};ll_{1}}\tilde{c}_{m_{1}}^{\dagger}\tilde{c}_{l_{2}}\tilde{c}_{m_{1}}\tilde{c}_{l_{2}}^{\dagger}\tilde{c}_{l_{1}}^{\dagger}\tilde{c}_{l_{1}}\\
	+&\frac{1}{2!}\sum_{m_{1}m_{2}m_{1}'}(\omega-E_{m_{1}m_{2};m_{1'}}+i0^{+})^{-1}A_{m_{1}m_{2};lm_{1}'}^{*}A_{m_{1}m_{2};lm_{1}'}\tilde{c}_{m_{1}'}^{\dagger}\tilde{c}_{m_{2}}\tilde{c}_{m_{1}}\tilde{c}_{m_{1}}^{\dagger}\tilde{c}_{m_{2}}^{\dagger}\tilde{c}_{m_{1}'}\\
	+&\cdots\\
	+&\tilde{c}_{l}^{\dagger}(\omega-E_{l}-i0^{+})^{-1}\left[1+\sum_{l_{1}}A_{ll_{1};ll_{1}}\tilde{c}_{l_{1}}^{\dagger}\tilde{c}_{l_{1}}+\sum_{m_{1}}A_{lm_{1};lm_{1}}^{*}\tilde{c}_{m_{1}}^{\dagger}\tilde{c}_{m_{1}}+\sum_{m_{1}l_{1}}A_{lm_{1};m_{1}l}^{*}A_{l_{1}l;ll_{1}}\tilde{c}_{m_{1}}^{\dagger}\tilde{c}_{m_{1}}\tilde{c}_{l_{1}}^{\dagger}\tilde{c}_{l_{1}}+\cdots\right]\tilde{c}_{l}\\
	+&\sum_{m_{1}l_{1}}\sum_{l_{2}\neq l}\tilde{c}_{l_{2}}^{\dagger}\tilde{c}_{l_{1}}^{\dagger}\tilde{c}_{l_{1}}(\omega-E_{l_{2}}-i0^{+})^{-1}A_{lm_{1};m_{1}l_{2}}^{*}A_{l_{1}l_{2};ll_{1}}\tilde{c}_{m_{1}}^{\dagger}\tilde{c}_{l_{2}}\tilde{c}_{m_{1}}\\
	+&\frac{1}{2!}\sum_{m_{1}m_{2}m_{1}'}\tilde{c}_{m_{1}}^{\dagger}\tilde{c}_{m_{2}}^{\dagger}\tilde{c}_{m_{1}'}(\omega-E_{m_{1}m_{2};m_{1'}}-i0^{+})^{-1}A_{m_{1}m_{2};lm_{1}'}^{*}A_{m_{1}m_{2};lm_{1}'}\tilde{c}_{m_{1}'}^{\dagger}\tilde{c}_{m_{2}}\tilde{c}_{m_{1}}\\
	+&\cdots.
\end{align*}
The first and fifth lines give the energy of the quasi-particle peaks with renormalized weight determined by the operators inside the square bracket, while the rest involving summation over poles at various energies give the continuum and peak broadening.

\section{2. Particle conservation in systems with spontaneously broken symmetry}

\subsection{a. Apparent ``exception'' with spontaneously broken symmetry?}

Figure 1 of the main text makes clear that internal \textit{dynamical} processes of number conserving systems cannot spontaneously break number conservation.
This may appear incompatible with the vast amount of literature employing the concept of spontaneously broken U(1) symmetry, which can be mapped to an effective number conservation.
Using two representative example as demonstration, this section will clarify the common confusion as a result of omitting the environmental influence during quantum state thermalization, explicit incorporation of which would confirm all the essential conclusions of the main text.

As a representative example, consider the Heisenberg model,
\begin{equation}
    H=-J\sum_{\langle ii^\prime \rangle}\mathbf{S}_i \cdot \mathbf{S}_{i^\prime}
    \label{Heisenberg}
\end{equation}
with ferromagnetic couplings, $J>0$, between nearest neighboring sites $i$ and $i^\prime$ in a 3D square lattice.
It is straightforward to verify that the fully polarized ferromagnetic states along any direction span the degenerate subspace of the ground state.
Given the continuous azimuthal degrees of freedom, it is natural to expect a convenient \textit{mathematical} mapping of such freedom to that of a U(1) phase, as part of the SO(3) group.

Indeed, representing states via quantization along the $z$-axis and choosing the fully up-spin polarized state,$\ket{I_{z}}$ as the vacuum for bosonic spin-lowering operator, $a_i$, at site $i$ for the Holstein-Primakoff transformation~\cite{holsteinfield1940},
\begin{equation}
    S_i^+= \sqrt{2S-n_i}a_i,~~
    S_i^-= a_i^\dagger\sqrt{2S-n_i},~~
    S_i^z= S-n_j,~~\text{with } n_i\equiv a^\dagger_i a_i,
\label{HP_transformation}
\end{equation}
a fully  polarized state along the direction $\mathbf{n}$ would read,
\begin{equation}
    \ket{I_\mathbf{n}}=\prod_i \{exp[-i\theta\frac{\sqrt{2S}}{2}(e^{i\phi}a_i^{\dagger}+e^{-i\phi}a_i)]\}\ket{I_{z}}
    \label{I_n}
\end{equation}
with $\theta$ and $\phi$ denoting the elevation and azimuth angels of the polarized direction.
One thus can conveniently regard an ordered state, say $\ket{I_x}$ with $\theta=\pi/2$ and $\phi=\pi/2$, as one with spontaneously broken U(1) symmetry (that picks a particular $\phi$) in such a mathematical expression.
Correspondingly, the total number $\sum_i n_i$ simultaneously fluctuates for $\theta\ne 0$ when the azimuthal direction matters.

Notice that in this representation,
\begin{equation}
    H= -J\sum_{<ii'>}\left[(S-n_i)(S-n_{i'})+\frac{1}{2}S(a^\dagger_ia_{i'}+a^\dagger_{i'}a_i)\right].
    \label{Heisenberg_in_a}
\end{equation}
still respects number conserving of $a^\dagger$, reflecting the conservation of total $S^z$ of the system.
And yet, the physical spontaneously broken symmetry state $\ket{I_x}$ contains \textit{coherent} number fluctuation and breaks the U(1) symmetry of $\phi$.
This \textit{appears} to provide a ``counter example'' against the statements related to Fig.1 in the main text.

\subsection{b. proper consideration on quantum state thermalization}

However, the above consideration misses the essential environmental influence on quantum state thermalization.
Specifically, the number conserving form of Eq.~\ref{Heisenberg_in_a} dictates that no matter how long the time scale is, dynamics described in $H$ can never modify the total number of $n_i$ or equivalently the total $S^z$.
In other words, there is no internal dynamical processes for the reference state $\ket{I_z}$ to thermalize into the target $\ket{I_x}$.

In the actual laboratory, ferromagnetic states emerge because of the preference of the external environment.
Specifically, within the long time scale of internal dynamics for quantum state thermalization, there must exist in average an effective directional preference that dictates the direction of quantization for all the quantum states.
The existence of such environmental influence, no matter how small, is essential in determining the direction for the preferred state.
Since its role is only to single out the $x$-direction with negligible strength, it suffices to represent its effect by a simple
\begin{equation}
H_{env}=-\epsilon S^x \propto -\frac{\epsilon}{2} (a^\dagger + a)
\end{equation}
with tiny $\epsilon$\cite{leggett2006quantum}.
Once including this essential $H_{env}$ as part of the full Hamiltonian $H$, one sees that $H$ no longer has number conservation, and over the long time period of quantum state thermalization, the system would slowly evolve into $\ket{I_x}$ even starting from the reference state $\ket{I_z}$.

Correspondingly, since $H_{env}$ does not respect the number conservation of total $n_i$(or conservation $S^z$), the ``violation'' of number conservation in the resulting $\ket{I_x}$ (c.f. Eq.~\ref{I_n}) is perfectly expected.
In other words, systems with spontaneously broken symmetry is \textit{not} exceptional to the mathematically rigorous consideration of number conservation discussed in the main text.

\subsection{c. Eigen-particle picture and number conservation of one-body quasi-particles}

Since eigen-particles absorb all the physical dynamics, their emergence at each energy/time scale perfectly simulates the dynamical processes for the quantum state thermalization of the same scale.
Therefore, for any quantum thermalized state, even with spontaneously broken symmetry, there \textit{must} exist a physical set of eigen-particles naturally consistent with the thermalized state.
For the above example, the presence of $H_{env}$ of long time scale dictates that the spins be quantized along the $x$-direction for long time scale dynamics, such that $\ket{I_x}$ naturally becomes the reference state and the raising and lowering operators in Eq.~\ref{HP_transformation} refer to those of $S^x$.
The full $H$, including $H_{env}$ now reads,
\begin{equation}
    H= -J\sum_{<ii'>}\left[(S-n_i)(S-n_{i'})+\frac{1}{2}S(a^\dagger_ia_{i'}+a^\dagger_{i'}a_i)\right]-\epsilon (S-n_i),
    \label{Heisenberg_in_n_x}
\end{equation}
with strict number conservation even with $H_{env}$ included.

Therefore, \textit{all} the implications of number conservation discussed in the main text immediately apply.
For example, all the eigenstates would have a well-defined total $n_i$, or equivalently total $S^x$.
Similarly, the one-body quasi-particles, being eigen-particles $\tilde{a}^\dagger_i$, carry exactly one quanta of the bare $a^\dagger$, or equivalently a $-\hbar$ for $S^x$, as expected from the standard magnon excitation.

\subsection{d. An example of number non-conserving systems}

Another representative example is the XXZ model,
\begin{equation}
    H_{XXZ}=-J\sum_{\langle ii^\prime \rangle}(S_i^x S_{i^\prime}^x + S_i^y S_{i^\prime}^y +\Delta S_i^z S_{i^\prime}^z)
    \label{XXZ}
\end{equation}
in the in-plane polarization regime, $|\Delta| < 1$, with ferromagnetic couplings, $J>0$, between nearest neighboring sites $i$ and $i^\prime$ in a 3D square lattice.
It has been well-established~\cite{andersonconcepts1997} that an in-plane ferromagnetic order emerges in the thermodynamic limit.
The $\ket{I_x}$ discussed above in the Heisenberg model again is one of the states with spontaneously broken symmetry.

Including the corresponding environmental influence $H_{env}=-\epsilon S^x$ and employ similar $S^x$-quantized eigen-representation as in the Heisenberg model, the Hamiltonian reads,
\begin{equation}
    H_{XXZ}= -J\sum_{<ii'>}\left[(S-n_i)(S-n_{i'})+\frac{1+\Delta}{4}S(a^\dagger_ia_{i'}+a^\dagger_{i'}a_i)+\frac{1-\Delta}{4}S(a^\dagger_ia^\dagger_{i'}+a_{i'}a_i)\right]-\epsilon(S-n_i).
    \label{XXZ_in_a_x_quantization}
\end{equation}
One again see that there always exists a set of eigen-particle representation consistent with the physical quantum state thermalization.
In this case, since the Hamiltonian no longer conserves particle number, the corresponding eigenstates and the one-body quasi-particles (magnon) no longer respect conservation of total $n_i$ (or equivalently total $S^x$.
These results are again perfectly consistent with our main conclusions.

\subsection{e. Lack of external preference to coherently break number conservation}

The above examples illustrate that systems with rotational symmetry can indeed spontaneously break the azimuth angular degree of freedom, equivalent to breaking the corresponding U(1) symmetry.
Physically such states can only thermalize under the influence of environment preference that needs to be included for a proper description of physical quantum state thermalization.

For the specific case of superconductivity, however, no physical external influence in nature is capable of \textit{coherently} break the \textit{electron} particle conservation.
Therefore, the number conservation in superconductivity must be strictly respected.

\section{3. Proper and improper Bogoliubov quasi-particles}
In this section we first illustrate dominant fluctuation in proper (number-conserving) Bogoliubov quasi-particles in BCS-like theories.
We then demonstrate how adapting an improper (number-non-conserving) mean-field can result in a familiar form of improper Bogoliubov quasi-particles.

Following BCS's original argument on s-wave pairing, for a local on-site interaction,
\begin{equation}
-\frac{V}{\Omega}\sum_{kpq}c_{k+q\uparrow}^{\dagger}c_{p-q\downarrow}^{\dagger}c_{p\downarrow}c_{k\uparrow},
\end{equation}
we consider only the dynamical effect of pairing fluctuation, namely,
\begin{equation}
-\frac{V}{\Omega}\sum_{kq}c_{k+q\uparrow}^{\dagger}c_{-k-q\downarrow}^{\dagger}c_{-k\downarrow}c_{k\uparrow},
\end{equation}
as in the BCS effective Hamiltonian, where $\Omega$ denotes the system size.

\subsection{a. Proper Bogoliubov quasi-particles}

To find the dressed particles that approximately diagonalize the Hamiltonian (as approximate eigen-particles), one can apply a canonical transformation via $\mathcal{U}=e^A$, with
\begin{equation}
    A=\frac{1}{\Omega}\sum_{1,2}\text{\ensuremath{\alpha_{12}}}c_{1\uparrow}^{\dagger}c_{-1\downarrow}^{\dagger}c_{-2\downarrow}c_{2\uparrow},
\end{equation}
where the matrix $\alpha$ is real and anti-symmetric ($\alpha_{12}=-\alpha_{21}$) and system-size insensitive.
(Further diagonalization for the rest less relevant fluctuation would resulting smaller corrections beyond this discussion.)
The resulting dressed particles (using $c^\dagger_{k\uparrow}$ as a simple illustration),
\begin{equation}
    \tilde{c}_{k\uparrow}^{\dagger}\equiv e^{-A}c_{k\uparrow}^{\dagger}e^{A}=c_{k\uparrow}^{\dagger}+[c_{k\uparrow}^{\dagger},A]+\frac{1}{2!}[[c_{k\uparrow}^{\dagger},A],A]+\cdots,
\end{equation}
now absorb the dominant fluctuations, including
\begin{align*}
[c_{k\uparrow}^{\dagger},A] & =[c_{k\uparrow}^{\dagger},\frac{1}{\Omega}\sum_{1,2}\text{\ensuremath{\alpha_{12}}}c_{1\uparrow}^{\dagger}c_{-1\downarrow}^{\dagger}c_{-2\downarrow}c_{2\uparrow}]\\
 & =-\frac{1}{\Omega}\sum_{1}\ensuremath{\alpha_{1k}}c_{1\uparrow}^{\dagger}c_{-1\downarrow}^{\dagger}c_{-k\downarrow},
\end{align*}
and
\begin{align*}
[[c_{k\uparrow}^{\dagger},A],A] & =[-\frac{1}{\Omega}\sum_{1}\ensuremath{\alpha_{1k}}c_{1\uparrow}^{\dagger}c_{-1\downarrow}^{\dagger}c_{-k\downarrow},\frac{1}{\Omega}\sum_{23}\text{\ensuremath{\alpha_{23}}}c_{2\uparrow}^{\dagger}c_{-2\downarrow}^{\dagger}c_{-3\downarrow}c_{3\uparrow}]\\
 & =\frac{1}{\Omega^{2}}\sum_{13}\ensuremath{\alpha_{1k}}\ensuremath{\alpha_{k3}}c_{k\uparrow}^{\dagger}c_{1\uparrow}^{\dagger}c_{-1\downarrow}^{\dagger}c_{-3\downarrow}c_{3\uparrow}+\cdots,
\end{align*}
for example.

Focusing on the leading fluctuations, the dressed particles as approximate eigen-particles have well-defined structures,
\begin{align}
\textcolor{blue}{\tilde{c}_{k\uparrow}^{\dagger}} & =\textcolor{blue}{c_{k\uparrow}^{\dagger}} \nonumber\\
 & -\frac{1}{\Omega}\sum_{1}c_{1\uparrow}^{\dagger}c_{-1\downarrow}^{\dagger}(\ensuremath{\alpha_{1k}}+\cdots)\textcolor{red}{c_{-k\downarrow}} \nonumber\\
 & -\frac{1}{2!}\frac{1}{\Omega^{2}}\sum_{11'}\textcolor{blue}{c_{k\uparrow}^{\dagger}}c_{1\uparrow}^{\dagger}c_{-1\downarrow}^{\dagger}(\ensuremath{\alpha_{1k}}\ensuremath{\alpha_{1'k}}+\cdots)c_{-1'\downarrow}c_{1'\uparrow} \nonumber\\
 & +\frac{1}{3!}\frac{1}{\Omega^{3}}\sum_{121'}c_{1\uparrow}^{\dagger}c_{-1\downarrow}^{\dagger}c_{2\uparrow}^{\dagger}c_{-2\downarrow}^{\dagger}(\ensuremath{\alpha_{1k}}\ensuremath{\alpha_{2k}}\ensuremath{\alpha_{1'k}}+\cdots)c_{-1'\downarrow}c_{1'\uparrow}\textcolor{red}{c_{-k\downarrow}} +\cdots,
\label{proper_qp}
\end{align}
showing a clear pattern containing even-order terms that renormalize the electrons, and odd-order terms that fluctuate ``into'' holes of their time-reversal counterpart.

\subsection{b. Improper Bogoliubov quasi-particles}

\begin{figure}[ht]
    \begin{centering}
    \includegraphics[scale=0.42]{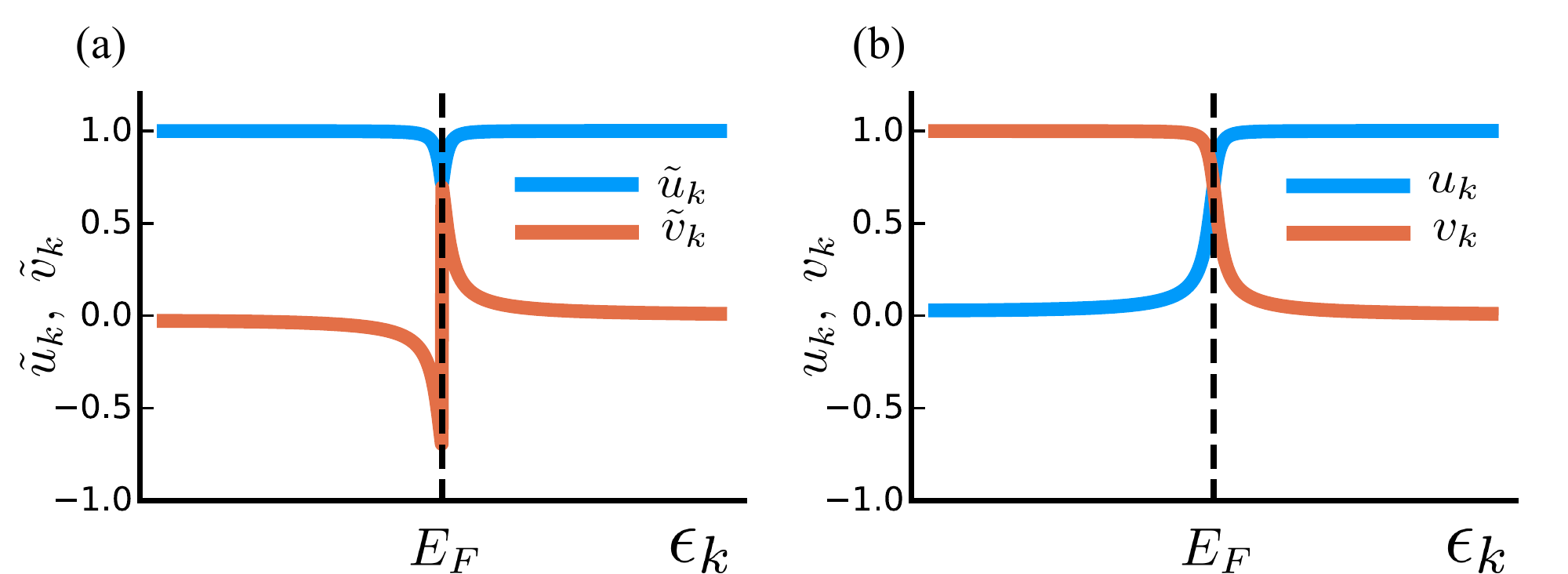}
    \par\end{centering}
    \caption{Sketches of (a) $\tilde{u}_k$ and $\tilde{v}_k$, and (b) $u_k$ and $v_k$, as a function of bare one-body energy $\epsilon_k$.
    Notice that in our picture, $|\tilde{v}_k|^2 < \frac{1}{2}$ always reflects the strength of number impairing fluctuations, while in the standard picture, $u_k$ and $v_k$ switch their roles in reflecting the fluctuation for $\epsilon_k \le E_F$ and $\epsilon_k > E_F$, respectively.
    Not accounting for this role switching obscures the previous attempts for repairing number conservation.}
    \label{pic:coherence_factor}
\vspace{-0.4cm}
\end{figure}

To illustrate that the standard (improper) Bogoliubov quasi-particles are improperly reduced eigen-particles, let's take the Jacobi ansatz $\ensuremath{\alpha_{1k}=\frac{1}{2}arctan(\frac{V}{\epsilon_{k}-\epsilon_{1}})}$. After adapting an improper mean-field that breaks number conservation,
\[
\Delta_0=-\frac{V}{\Omega}\sum_{1}\braket{c_{1\uparrow}^{\dagger}c_{-1\downarrow}^{\dagger}}.
\]
the proper Bogoliubov quasi-particles now reduce to the improper form,
\begin{align}
\textcolor{blue}{\tilde{c}_{k\uparrow}^{\dagger}}&\approx\tilde{u}_{k}\textcolor{blue}{c_{k\uparrow}^{\dagger}}-\tilde{v}_{k}\textcolor{red}{c_{-k\downarrow}}\\
\textcolor{blue}{\tilde{c}_{-k\downarrow}}&\approx\tilde{u}_{k}\textcolor{blue}{c_{-k\downarrow}}+\tilde{v}_{k}\textcolor{red}{c_{k\uparrow}^\dagger},
\end{align}
with $\tilde{u}_{k}\equiv cos(\phi) \ge \tilde{v}_{k}\equiv sin(\phi)$ and $\phi\equiv \frac{1}{2}arctan(\frac{\Delta_0}{\epsilon_k - E_F})$ preserving the canonical commutation relationship, $\{\tilde{c}_l,\tilde{c}^\dagger_{l^\prime}\}=\delta_{ll^\prime}$,
as shown in Fig.~\ref{pic:coherence_factor}(a).
Here the opposite signs of the relative phase results from the anti-symmetric property of $\alpha_{kk^\prime}$.

Now, concerning the superconducting ground states of fixed $N$ particles, $\ket{N}_0$, its quasi-particle excitations, as measured by one-body Green's function, would then include adding one more eigen-particle (active when the bare particle energy is above the Fermi energy, $\epsilon_k > E_F$),
\begin{equation}
\gamma^\dagger_{k\uparrow}=\textcolor{blue}{\tilde{c}^\dagger_{k\uparrow}}\approx\tilde{u}_{k}\textcolor{blue}{c_{k\uparrow}^{\dagger}}-\tilde{v}_{k}\textcolor{red}{c_{-k\downarrow}},
\label{adding}
\end{equation}
or remove an eigen-particle (adding one more hole, when $\epsilon_k \le E_F$),
\begin{equation}
\gamma^\dagger_{k\uparrow}=-\textcolor{blue}{\tilde{c}_{-k\downarrow}}\approx - (\tilde{u}_{k} \textcolor{blue}{c_{-k\downarrow}}+\tilde{v}_{k}\textcolor{red}{c^\dagger_{k\uparrow}}),
\label{removing}
\end{equation}
or equivalently one arrives at the familiar form of improper Bogoliubov quasi-particles~\cite{schrieffertheory2018},
\begin{align}
\gamma_{k\uparrow}^{\dagger}&\approx u_{k}c_{k\uparrow}^{\dagger}-v_{k}c_{-k\downarrow}
\label{std_BG}
\end{align}
by choosing
\begin{align}
u_k&\equiv \tilde{u}_k, ~~~~v_k\equiv \tilde{v}_k  \mathrm{;}~~~~\epsilon_k > E_F \nonumber \\
u_k&\equiv -\tilde{v}_k, ~v_k\equiv \tilde{u}_k\mathrm{;}~~~~ \epsilon_k \le E_F
\end{align}
as shown in Fig.~\ref{pic:coherence_factor}(b).
That is, the standard improper Bogoliubov quasi-particles are indeed reduced eigen-particles under an improper (number-non-conserving) mean-field.

\subsection{c. An error in previous attempts toward proper Bogoliubov quasi-particles}

Recall that one-body quasi-particle excitations contain probabilities of both removing and adding a particle to the system, as can be probed separately by photoemission and inverse photoemission experiments.
Therefore, a straightforward patch-up~\cite{blonderTransition1982, tinkhamintron2004, lin2022some} of the improper Bogoliubove quasi-particles via,
\begin{equation}
\gamma^\dagger_{k\uparrow}\approx u_{k}c_{k\uparrow}^{\dagger}-v_{k} C^\dagger_k c_{-k\downarrow},
\label{improper_fix}
\end{equation}
is reasonable for the particle addition processes ($\epsilon_k > E_F$), with $C^\dagger_k$, defined as $\frac{1}{\Omega}\sum_{1}c^\dagger_{1\uparrow}c_{-1\downarrow}^{\dagger}\alpha_{1k}$, properly compensating the annihilation operator.
However, the same patch-up seems highly questionable in the particle removal processes ($\epsilon_k \le E_F$), as it \textit{adds} exactly a particle instead.

Indeed, Eq.~\ref{adding} and \ref{removing} make it clear that these two processes need to be distinguished in remedying the improper Boboliubov quasi-particles.
According to Eq.~\ref{proper_qp}, including only the lowest order fluctuation in $\tilde{c}^\dagger_{k\uparrow}$, Eq.~\ref{adding} and \ref{removing} would give,
\begin{align}
\label{apprx_adding}
\gamma^\dagger_{k\uparrow}&=\textcolor{blue}{\tilde{c}^\dagger_{k\uparrow}}\approx\tilde{u}_{k}\textcolor{blue}{c_{k\uparrow}^{\dagger}}-\tilde{v}_{k} C^\dagger_k \textcolor{red}{c_{-k\downarrow}};~~~~~~~~~~~~~~~~~~~\epsilon_k > E_F\\
\gamma^\dagger_{k\uparrow}&=-\textcolor{blue}{\tilde{c}_{-k\downarrow}}\approx -(\tilde{u}_{k} \textcolor{blue}{c_{-k\downarrow}} + \tilde{v}_{k} \textcolor{red}{c^\dagger_{k\uparrow}} C_{k});~~~~~~~~\epsilon_k \le E_F,
\label{apprx_removing}
\end{align}
respectively.
While the previous attempts via Eq.~\ref{improper_fix} are consistent with Eq.~\ref{apprx_adding} for the particle additional processes, it seriously contradicts Eq.~\ref{apprx_removing} for particle removal processes.

As a matter of fact, this serious inconsistency clearly manifests itself through the relative strengths of $u_k$ and $v_k$ in Eq.~\ref{std_BG}.
As shown in Fig.~\ref{pic:coherence_factor}, for the particle removal processes ($\epsilon_k \le E_F$), $u_k$ is actually weaker than $v_k$ in strength.
In other words, in this case, $c_{k\uparrow}^{\dagger}$ is the fluctuation that needs to be patched up (by $C_{k}$) to match the dominant removal process, $c_{-k\downarrow}$, as shown by the red component in Eq.~\ref{apprx_removing}.
It is quite curious that such an obvious inconsistency has not been noticed in the literature.

\end{document}